\documentclass[aps,prd,twocolumn,showpacs,amsmath,amssymb,nofootinbib,superscriptaddress]{revtex4-2}
\usepackage{graphicx}  
\usepackage{color}
\usepackage{times}
\usepackage[utf8]{inputenc} 
\usepackage{float}
\usepackage{bm}
\usepackage{ulem}
\usepackage{multirow}
\usepackage{makecell}
\usepackage{url}
\usepackage{natbib}
\usepackage{mathrsfs}
\usepackage{physics}
\usepackage{comment}
\usepackage[table,xcdraw]{xcolor}
\usepackage{booktabs,makecell}
\usepackage{amsmath}
\usepackage{pifont}
\usepackage[colorlinks=true,citecolor=blue,urlcolor=blue,linkcolor=blue]{hyperref}
\usepackage{microtype}
\usepackage[none]{hyphenat}

\DeclareUnicodeCharacter{03BB}{\ensuremath{\lambda}} 
\DeclareUnicodeCharacter{039B}{\ensuremath{\Lambda}} 

\begin{document}

\title{Constraints on the parameter space in dark matter admixed neutron stars}

\author{Ankit Kumar}
\email{ankit.kumar@riken.jp}
\affiliation{Astrophysical Big Bang Laboratory, RIKEN, Saitama 351-0198, Japan}

\author{Hajime Sotani}
\affiliation{Astrophysical Big Bang Laboratory, RIKEN, Saitama 351-0198, Japan}
\affiliation{Interdisciplinary Theoretical \& Mathematical Science Program (iTHEMS), RIKEN, Saitama 351-0198, Japan}

\date{\today}

\begin{abstract}
We investigate the impact of dark matter on neutron star properties using the relativistic mean-field theory. By incorporating the dark matter model, we explore how dark matter parameters, specifically dark matter mass and Fermi momentum, influence nuclear saturation properties, the equation of state, and the mass-radius relationship of neutron stars. We also examine the universal relation between dimensionless tidal deformability and compactness in the presence of dark matter. Our results show that the inclusion of dark matter significantly alters nuclear saturation properties, leading to higher incompressibility and symmetry energy values. Notably, higher dark matter Fermi momenta and masses result in more compact neutron star configurations with reduced radii and lower maximum masses, highlighting a complex interplay between dark matter and nuclear matter. Deviations from the universal relation are observed with dark matter inclusion, particularly for neutron stars with lower compactness. By leveraging observational data from PSR J0740+6620, GW170817, and Neutron star Interior Composition Explorer (NICER) measurements of PSR J0030+0451, we derive stringent constraints on dark matter parameter space within neutron stars, emphasizing the necessity of integrating multimodal observations to delineate the properties of dark matter along with neutron stars. Our findings underscore the importance of considering dark matter effects in neutron star modeling and suggest potential refinements for current theoretical frameworks to accurately predict neutron star properties under various astrophysical conditions.
\end{abstract}

%

\maketitle

\section{Introduction}
\label{sec:1}

Neutron stars (NSs), the remnants of massive stellar cores, present a unique opportunity to explore the fundamental properties of matter under extreme conditions. These compact objects, characterized by densities exceeding those of atomic nuclei, exhibit a variety of intriguing phenomena including rapid rotation, intense magnetic fields, and emission across the electromagnetic spectrum. Theoretical models suggest that the immense pressures within NSs counterbalance gravitational forces that would otherwise lead to collapse, thereby preventing the formation of black holes. Observationally, NSs have been extensively studied through their x-ray, gamma-ray, and radio emissions, with pulsars-rapidly rotating NSs emitting beams of electromagnetic radiation—providing critical insights into their internal structure and gravitational fields. The precise timing of pulsar signals has provided valuable data for testing theories of gravity and understanding the dynamics of NSs. Additionally, the advent of gravitational wave (GW) astronomy has opened new frontiers in the study of NSs. Events such as GW170817 \cite{PhysRevLett.119.161101} have demonstrated the potential of gravitational waves to reveal insights into NS mergers and inspirals. Observations of X-ray oscillations from millisecond pulsars and the shape of their pulse profiles provide additional data for imposing constraints on NS models \cite{annurev:/content/journals/10.1146/annurev-astro-040312-132617}. Continuous gravitational wave searches, targeting phenomena like elastic deformations and unstable oscillation modes, offer promising avenues for further exploration of NS properties. These multi-messenger observations are crucial for validating theoretical models and enhancing our understanding of the most enigmatic and extreme states of matter.

Dark matter (DM), which constitutes approximately 27\% of the Universe's mass-energy content, yet remains one of the most significant mysteries in modern astrophysics and cosmology. The existence of DM, initially called “missing mass”, was first suggested by Zwicky in 1933 \cite{Zwicky:1933gu, 2017arXiv171101693A} and later supported by Rubin and Ford's optical studies of galaxies in the 1970s \cite{1970ApJ...159..379R, 1978ApJ...225L.107R}. Various models have been proposed to explain DM, including weakly interacting massive particles and asymmetric DM. While traditional models have often assumed DM to be neutral, recent studies have explored the possibility of charged massive particles and their implications for astrophysical observations. These models suggest that DM could form either a dense core inside NSs or an extended halo surrounding them, each having distinct effects on the NS's properties. In asymmetric DM models, where DM particles and their antiparticles have different properties, such as different masses or abundances, and this asymmetry leads to a net excess of DM particles over antiparticles \cite{2014PhR...537...91Z}; over time DM particles could accumulate in the cores of NSs, potentially leading to their collapse into black holes. This accumulation is influenced by the DM's mass and self-interaction strength, which determine whether it forms a dark core or a halo \cite{PhysRevLett.113.191301}. For instance, light DM particles tend to form halos, while heavier ones prefer cores \cite{PhysRevD.84.107301}. Despite extensive efforts, its nature and properties are still largely unknown due to its weak interactions with ordinary matter. The detection of gravitational waves by the LIGO and Virgo Collaborations, particularly from NS mergers like GW170817, has opened new avenues for probing the nature of DM. This observation provides constraints on DM properties by analyzing the power spectral density of GW emissions, especially if DM cores are present within NSs \cite{ELLIS2018607}. 
The possibility of the 2.6 $M_{\odot}$ compact object in the binary merger GW190814 \cite{Abbott_2020} being a DM admixed NS has also been investigated, highlighting the potential of gravitational wave astronomy in studying DM \cite{PhysRevD.104.063028}. The presence of DM can significantly alter the mass-radius relationship, surface temperature, and tidal deformability of NSs, as well as their gravitational wave signatures \cite{PhysRevD.107.083037}. As our observational capabilities improve, particularly with multimessenger astronomy, the study of DM admixed NSs promises to yield valuable insights into the fundamental properties of DM and its role in the cosmos. The combination of gravitational wave data, x-ray observations, and theoretical models will continue to enhance our understanding of these enigmatic particles and their interactions with one of the universe's most extreme environments.

The concept of DM admixed NSs has been explored in various studies, with neutralinos often considered as fermionic DM candidates. Neutralinos, predicted by supersymmetric extensions of the Standard Model, are stable, weakly interacting massive particles that could significantly influence the properties of NSs. Including neutralinos in NS models allows researchers to theoretically study the impact of DM on the equation of state (EOS) and other observable characteristics of NSs. For instance, Bertone and Fairbairn (2008) and Kouvaris (2008) explored how neutralinos could accumulate in NSs, affecting their thermal evolution and potentially leading to observable signatures in the form of surface temperature variations \cite{PhysRevD.77.043515, PhysRevD.77.023006}. Additionally, Kouvaris and Tinyakov (2010) and P\'{e}rez-Garc\'{i}a and Silk (2012) investigated the kinematics and rotation properties of NSs influenced by neutralino DM \cite{PhysRevD.82.063531, PhysRevLett.105.141101}. The presence of neutralinos within NSs can alter the mass-radius relationship, as well as the tidal deformability, which is crucial for interpreting gravitational wave signals from NS mergers. Recent studies by Ellis et al. (2018) and Nelson, Reddy, and Zhou (2019) have shown that the inclusion of a DM component, such as neutralinos, can significantly modify the tidal deformability parameter, thereby providing constraints on the DM properties from gravitational wave observations \cite{PhysRevD.97.123007, Nelson_2019}. Furthermore, the interaction between neutralinos and nucleons, mediated by Higgs bosons, has been studied to understand its impact on the EOS of NSs. This interaction can lead to the formation of either a DM core or halo, depending on the properties of the neutralinos, such as their mass and self-interaction strength \cite{PhysRevD.105.023001}. In addition to these studies, recent research has investigated the structure of NSs admixed with fermionic self-interacting DM. These studies analyze different scenarios based on the properties of dark fermions, interaction mediators, and self-interacting strengths, providing constraints on DM model parameters using multimessenger astronomy \cite{10.1093/mnras/stad3658}. 

The model adopted in the present work to put constraints on the mass-momentum space of DM admixed NS with neutralinos as fermionic DM candidates has its roots in this article \cite{PhysRevD.96.083004} by Panotopoulos and Lopes. They applied relativistic mean-field (RMF) theory to NSs, assuming that fermionic DM particles are trapped inside the star. They computed the EOS for both baryonic matter and DM, showing how the presence of DM could alter the mass-radius relationship and the maximum mass of NSs. Later, this approach was explored further in various studies, that confronted the EOS with gravitational wave constraints from the GW170817 event. Using the RMF model with NL3 parametrization, Das, Malik, and Nayak demonstrated that the inclusion of DM softens the EOS, lowering the tidal deformability and bringing the model into an agreement with observational data \cite{PhysRevD.99.043016}. Building on these foundational studies, various studies further investigate the effects of fermionic DM on inspiral properties of NS, including the mass-radius relationship, a moment of inertia, tidal Love numbers, curvature, and thermal relaxation \cite{10.1093/mnras/staa1435, Das_2021, 10.1093/mnras/stab2387, 10.1093/mnras/stac1013}. In all these works, the model has been adopted to study NSs with a fixed particle mass of DM candidate, and no tight constraints have been put forward on the parameter space of DM. In addition, the crust region of a NS has not been taken into account properly in most of the previous studies. Thus, in this study, by considering the crust region together with the NS core and assuming that DM exists only inside the NS core, we systematically explore DM admixed NS models to constrain the parameter space of DM particle's mass and Fermi momentum surface in light of the astronomical constraints, i.e., current gravitational wave constraints, particularly GW170817, the mass observation of the massive NS, and the constraint on the NS mass and radius via the Neutron star Interior Composition Explorer (NICER). Additionally, we tested the universal relations, i.e., the relation between the dimensionless tidal deformability and stellar compactness, for different DM admixture cases to assess the applicability of the current DM admixed NS model. Our findings aim to provide a comprehensive understanding of the interplay between DM properties and NS observables, offering new insights into the constraints on RMF parameter sets imposed by GW events. 

The structure of this paper is designed to methodically analyze the effects of DM within NS models, guided by the RMF theory. Section \ref{sec:2} introduces the fundamental formalism of RMF alongside DM interactions, setting the groundwork for the study of DM-admixed NSs. In Sec.~\ref{sec:3a}, we investigate how DM influences the saturation parameters of nuclear matter, highlighting changes in properties like binding energy and incompressibility. Section \ref{sec:3b} delves into the impact of DM on key NS observables, such as mass-radius profiles and tidal deformability, and establishes constraints on DM parameter space based on recent astronomical data. In Sec.~\ref{sec:3c}, we evaluate the universality of the $\Lambda-M/R$ relationship in the presence of DM, where $\Lambda$ denotes the NS dimensionless tidal deformability and $M/R$ is the compactness of the NS, testing the robustness of this relation under varied DM influences. The paper concludes with Sec.~\ref{sec:4}, where we summarize our findings and discuss the implications of DM within NSs, contributing to the broader understanding of these compact objects in astrophysics.

\section{DM admixed NS model}
\label{sec:2}

The RMF theory is a widely used theoretical framework for modeling the interactions between nucleons in NSs. The RMF model effectively captures the essential features of nuclear interactions by incorporating the meson fields, which generate mean-field potentials felt by the nucleons, providing a robust framework for studying dense nuclear matter \cite{WALECKA1974491, Serot:1991st, 1996csnp.book.....G}. In the RMF approach, nucleons interact through the exchange of mesons, which mediate the strong nuclear force and generate mean-field potentials that nucleons experience, allowing for a self-consistent calculation of the EOS of nuclear matter. Over time, the RMF Lagrangian has evolved to incorporate more sophisticated interactions to better capture the complexities of nuclear matter. Modern RMF parameter sets now include self-coupling and cross-coupling terms of the mesons, extending up to fourth order \cite{1977NuPhA.292..413B, PhysRevC.55.540, PhysRevC.55.540, FURNSTAHL1997441, PhysRevC.85.024302}. These additional terms allow for a more accurate representation of the nonlinear interactions within the nuclear medium, enhancing the predictive power of the RMF model. The RMF Lagrangian density typically includes contributions from isoscalar scalar mesons ($\sigma$), isoscalar vector mesons ($\omega$), isovector vector mesons ($\rho$), and isovector scalar mesons ($\delta$). The general form of the currently used Lagrangian can be expressed as \cite{Singh_2014, Kumar2020}:
\begin{widetext}
\begin{eqnarray}\label{lag}
{\cal L}_{\rm NM} & = &  \sum_{i=p,n} \bar\psi_{i}
\left\{\gamma_{\mu}(i\partial^{\mu}-g_{\omega}\omega^{\mu}-\frac{1}{2}g_{\rho}\vec{\tau}_{i}\!\cdot\!\vec{\rho}^{\,\mu})
-(M_n-g_{\sigma}\sigma-g_{\delta}\vec{\tau}_{i} \!\cdot\!\vec{\delta})\right\}\psi_{i}
+\frac{1}{2} \partial^{\mu}\sigma\,\partial_{\mu}\sigma
-\frac{1}{2}m_{\sigma}^{2}\sigma^2
+\frac{1}{2}m_{\omega}^{2}\omega^{\mu}\omega_{\mu}
\nonumber\\
&&
-\frac{1}{4}F^{\mu\nu}F_{\mu\nu}
-g_{\sigma}\frac{m_{\sigma}^2}{M_n} \left(\frac{\kappa_3}{3!} + \frac{\kappa_4}{4!}\frac{g_{\sigma}}{M}\sigma\right) \sigma^3
+\frac{\zeta_0}{4!}g_\omega^2 (\omega^{\mu}\omega_{\mu})^2
+\frac{1}{2}\frac{g_{\sigma}\sigma}{M_n}\left(\eta_1+\frac{\eta_2}{2} \frac{g_{\sigma}\sigma}{M_n}\right)m_\omega^2\omega^{\mu}\omega_{\mu}
+\frac{1}{2}m_{\rho}^{2}\rho^{\mu}\!\cdot\!\rho_{\mu} 
\nonumber\\
&& \null
-\frac{1}{4}\vec R^{\mu\nu}\!\cdot\!\vec R_{\mu\nu}
+\frac{1}{2}\eta_{\rho}\frac{m_{\rho}^2}{M_n}g_{\sigma}\sigma(\vec\rho^{\,\mu}\!\cdot\!\vec\rho_{\mu}) 
-\Lambda_{\omega}g_{\omega}^2g_{\rho}^2(\omega^{\mu}\omega_{\mu})(\vec\rho^{\,\mu}\!\cdot\!\vec\rho_{\mu})
+\frac{1}{2}\partial^{\mu}\vec\delta\,\partial_{\mu}\vec\delta-\frac{1}{2}m_{\delta}^{2}\vec\delta^{\,2} 
+\sum_{j=e^{-},\mu}\bar\phi_{j}(i\gamma_{\nu}\partial^{\nu}-m_{j})\phi_{j}, \nonumber \\ 
\end{eqnarray}
\end{widetext}
where $\psi_i$ represents the nucleon field, $\sigma$, $\omega$, $\vec{\rho^{\mu}}$, and $\vec{\delta}$ are the meson fields, and the terms $F^{\mu\nu}$ and $R^{\mu\nu}$ denote the field tensors for the $\omega$ and $\rho$ mesons, respectively. $M_n$ represents the mass of the nucleon, i.e., 939 MeV; $\tau_{i}$ stands for the isospin operator; $g_{k}$ ($k = \sigma$, $\omega$, $\rho$, and $\delta$) represents the coupling strengths for respective mesons; and the other quantities (i.e. $\kappa_3$, $\kappa_4$, $\zeta_0$, $\eta_1$, $\eta_2$, $\eta_\rho$, and $\Lambda_\omega$) represent self- and cross-coupling strengths among mesons. The last term in the Lagrangian represents the lepton contribution (i.e. electrons and muons), and their inclusion is essential for ensuring that the NS remains charge neutral and in a state of beta equilibrium, where the rates of beta decay and electron capture are balanced \cite{1983bhwd.book.....S}. Utilizing the Euler-Lagrange equations of motion, we obtain a set of coupled differential equations for the meson fields from the above Lagrangian. These equations are then solved self-consistently until they converge to a stable configuration. Once the meson field values are determined, they are substituted into the expressions for the energy density and pressure which can be easily obtained by calculating the components of energy-momentum tensor \cite{Kumar2020}. This process yields the EOS for NSs, where the corresponding RMF parameter sets with their specific coupling constants are used to accurately model the interactions within the dense matter. 

The coupling constants in the RMF Lagrangian, which determine the strength of the interactions between nucleons and mesons, are represented by different RMF parameter sets. These parameter sets are calibrated to reproduce mainly the properties of finite nuclei and symmetric nuclear matter. In the present work, we used six RMF parameter sets to span the nature of the EOS from the stiffest to the softest region, namely NL3 (oldest and stiffest), BigApple (recently developed, with a balanced description of finite nuclei and massive NS), IOPB-I (moderately stiff with fewer couplings), G3 (includes higher order couplings and $\delta$ meson), IU-FSU, and FSUGold (relatively soft EOS). The coupling constants and masses of mesons for the RMF parameter sets adopted in this study can be found in the corresponding references, i.e., NL3 \cite{PhysRevC.55.540}, BigApple \cite{PhysRevC.102.065805}, IOPB-I \cite{PhysRevC.97.045806}, G3 \cite{KUMAR2017197}, IU-FSU \cite{PhysRevC.82.055803}, and FSUGold \cite{PhysRevLett.95.122501}.
\begin{figure}[tbp]
    \centering
    \includegraphics[width=\columnwidth]{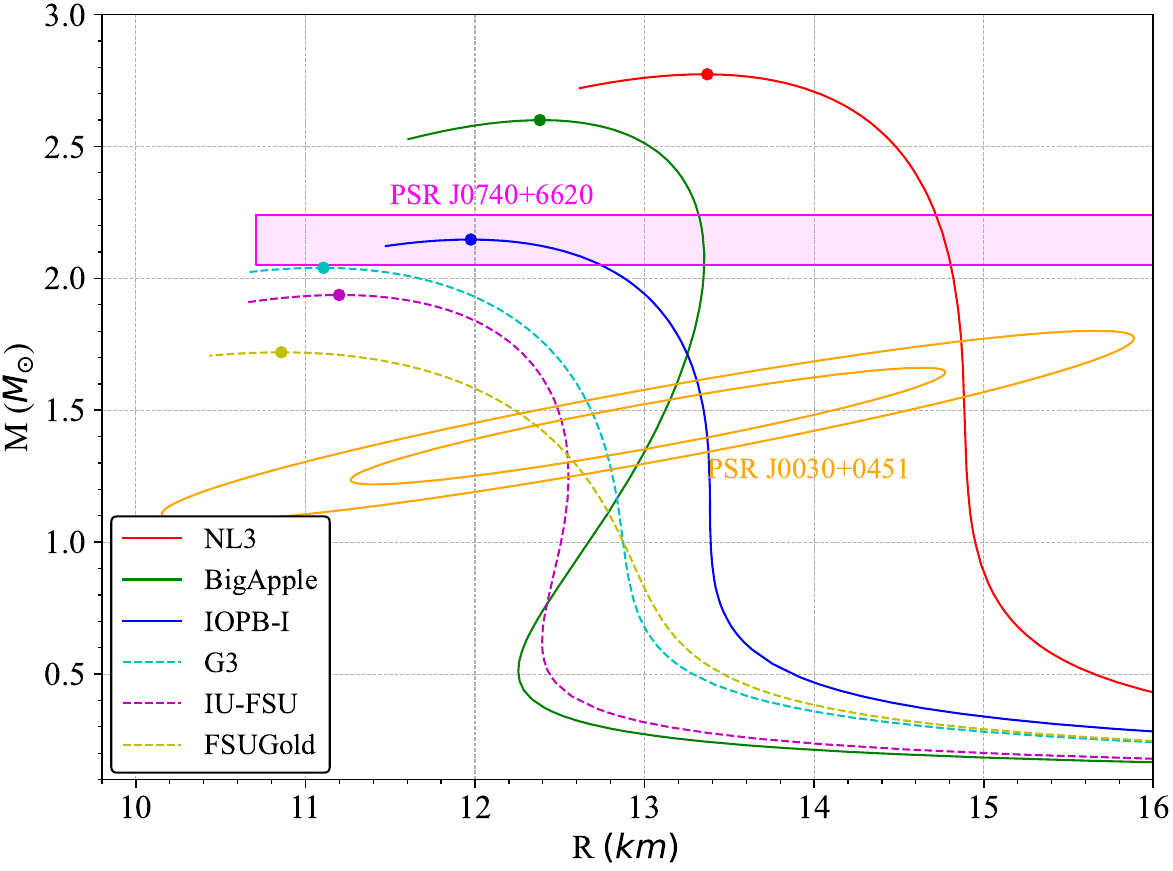} 
    \caption{NS mass and radius relations constructed with several EOSs adopted in this study without any effects of DM. The mark on each line denotes the stellar model with the maximum mass. For reference, two astronomical constraints are also shown, i.e., the mass of the massive NS, i.e., $M=2.08^{+0.07}_{-0.07}M_\odot$ (68.3\% credibility), and mass and radius region constrained from NICER observation for PSR J0030+0451.}
    \label{fig:figure1}
\end{figure}

For the crust part of the NS, we use the SLy4 data to create a comprehensive EOS that spans from the core to the crust. The SLy4 EOS is well-suited for describing the properties of the NS crust, providing a smooth transition to the core EOS generated by the considered RMF parameter sets \cite{CHABANAT1998231, refId0}. Now, the mass-radius ($M$-$R$) profile for a given EOS of NS matter is calculated by integrating the Tolman-Oppenheimer-Volkoff (TOV) equations \cite{PhysRev.55.364, PhysRev.55.374}, which are the relativistic equations of hydrostatic equilibrium, from the center of the star to its surface with appropriate boundary conditions. These equations take into account the balance between the gravitational force and the pressure gradient within a NS, and are given by:
\begin{eqnarray}
    \frac{dP(r)}{dr} &=& -\frac{ \left( \varepsilon(r) + P(r) \right) \left( m(r) + 4 \pi r^3 P(r) \right)}{r \left( r - 2 m(r) \right)}, \\
    \frac{dm(r)}{dr} &=& 4 \pi r^2 \varepsilon(r),
\end{eqnarray}
where $P(r)$ is the pressure, $\varepsilon(r)$ is the energy density, $m(r)$ is the enclosed gravitational mass, and $r$ is the radial position. The $M$-$R$ profiles for the EOS derived from the six RMF parameter sets (NL3, BigApple, IOPB-I, G3, IU-FSU, and FSUGold) without any DM admixture are depicted in Fig.~\ref{fig:figure1}. In the same figure, for reference, we also show the observational data: the mass of PSR J0740+6620, which is one of the massive NSs observed so far, i.e., $M=2.08^{+0.07}_{-0.07}M_\odot$ (68.3\% credibility) \cite{2020NatAs...4...72C, Fonseca_2021} \footnote{We note that the theoretical maximum mass is given by $M_{\rm max}/M_\odot\approx 2.856 + 1.511 \times 10^{-3} (K_0 L^2)^{1/3}/(1\ {\rm MeV})$ with the nuclear saturation parameters, $K_0$ and $L$, (see Eq.~(\ref{eq:bulk}) for their definitions) \cite{PhysRevC.95.025802}.}, and the constraints on the NS mass and radius from the NICER analysis for PSR J0030+0451 \cite{Riley_2019, Miller_2019, universe6060081}. The NICER has also reported the radius of PSR J0740+6620, i.e., $R=12.39^{+2.63}_{-1.68}$ km (95\% credibility) \cite{Riley_2021} and $R=11.8$ -$13.4$ km (90\% credibility) \cite{Miller_2021}. In Fig.~\ref{fig:figure1}, we show the minimum radius of PSR J0740+6620, adopting the result in \cite{Riley_2021}, i.e., $R\ge 10.71$ km. In addition to these astronomical constraints on the NS mass and radius, the gravitational wave observations at the binary NS merger also endue us the NS dimensionless tidal deformability, which leads to the constraint on the $1.4M_\odot$ NS radius, i.e., it should be less than 13.6 km \cite{PhysRevLett.120.172703}. Considering this constraint, as mentioned earlier, NL3, being the RMF parameter set with the stiffest EOS \footnote{In this study, we simply refer the EOS to as stiffer (softer), when the maximum mass of a NS is larger (smaller).}, predicts a higher maximum mass for NS but results in too large a radius around a canonical NS ($1.4\, M_{\odot}$). BigApple, while also predicting high maximum masses, shows a better agreement with observational data for low-mass NSs, as its $M$-$R$ profile curve has lower radii for low-mass stars compared to NL3. The IOPB-I parameter set offers a balanced EOS that provides a maximum mass prediction that fits well within the observed mass range for PSR J0740+6620 and the $M$-$R$ space constraints from NICER data. On the other end of the spectrum, FSUGold, being the softest EOS considered, predicts the lowest maximum mass for NSs after solving the TOV equations. All these calculations for Fig. \ref{fig:figure1} are performed without any DM admixture, providing a baseline for the EOS of NSs purely based on nuclear interactions as described by the RMF theory. In the further part of this section, we will introduce the Lagrangian for DM interactions, and in the subsequent sections, we will explore how the inclusion of DM affects the observables of NS and derive constraints for mass-momentum space of DM from experimental data and astronomical observations.

Now, the Lagrangian to explore the DM interaction effects, considering fermionic DM candidate, with mass $M_{\chi}$, can be formulated as \cite{PhysRevD.96.083004}:
\begin{align}
    \mathcal{L}_{\rm DM} =& \,\bar\chi\left[i\gamma^\mu \partial_\mu - M_\chi + y h \right]\chi 
    \nonumber \\
    & + \frac{1}{2}\partial_\mu h \partial^\mu h - \frac{1}{2} M_h^2 h^2 + \frac{f\,M_{n}}{v} \bar \psi h \psi,   
    \label{eq:L_DM}
\end{align}
where $\chi$ expresses the field for the DM candidate; $h$ represents the Higgs field; $M_{\chi}$ is the mass of DM particle; $M_{h}$ is the mass of Higgs (=125 GeV); $y$ is the coupling for Higgs and DM (fixed at 0.06 in present calculations); and $fM_{n}/v$ represents the nucleon-Higgs coupling strength ($\approx 1.145 \times 10^{-3}$), where $v$ is the Higgs vacuum expectation value given by $v=246$ GeV, adopting $f=0.3$ \cite{PhysRevD.96.083004} . Using the same mathematical technique of energy-momentum tensor as mentioned earlier and mean-field approximation for Higgs field as well, the expressions of energy density and pressure for the DM part can be derived as:
\begin{align}
    \mathcal{E}_{\rm DM} &= \frac{2}{(2\pi)^{3}} \int_{0}^{k_f^{\rm DM}} d^3k \sqrt{k^2 + (M_{\chi}-yh)^2} + \frac{1}{2} M_{h}^{2} h^{2}, \nonumber \\
    P_{\rm DM} &= \frac{2}{3(2\pi)^{3}} \int_{0}^{k_f^{\rm DM}} \frac{k \,d^3k}{\sqrt{k^2 + (M_{\chi}-yh)^2}} - \frac{1}{2} M_{h}^{2} h^{2},
\end{align}
where the value of DM Fermi momenta, $k_f^{\rm DM}$, must be selected to ascertain the contribution of DM. We note that the DM contribution completely disappears by definition if one sets $k_f^{\rm DM}=0$. In practice, we examine the nuclear properties and NS models by changing $M_\chi$ and $k_f^{\rm DM}$ in this study. This is because the dependence on the other parameters, such as $y$ and $fM_n/v$ in Eq.~(\ref{eq:L_DM}), is almost negligible in the DM model we adopted here. For instance, for a specific DM admixed NS configuration, using the IOPB-I parameter set with $M_\chi=100$ GeV and $k_f^{\rm DM} = 0.04$ GeV, we confirm that a minimal variation of 0.14\% in maximum mass and a 3.42\% change in the corresponding radius with $y=$ 0.004 and 0.06, while the permissible range for parameter $y$ is 0.001 to 0.1 \cite{PhysRevD.96.083004}. On the other hand, the Higgs-nucleon coupling parameter, $f$, is expected in a small range of $0.341\pm0.021$ or $0.348\pm 0.015$~\cite{PhysRevD.88.055025}. Then, the total energy density ($\mathcal{E}$) for a DM admixed star can be defined as $\mathcal{E = E_{\rm NM} + E_{\rm DM}}$, and similarly the pressure from both the parts can be added to get the overall pressure of the system. It is worth mentioning that when we talk about a DM admixed nuclear matter in the next section then the $\mathcal{E_{\rm NM}}$ contains the contribution only from neutrons and protons and not from the leptons i.e. the last term of the Lagrangian in Eq.~(\ref{lag}) is not present in the definition of nuclear matter, it has been solely added to construct NS matter.

In this adopted model, DM particles interact indirectly with nucleons via the Standard Model Higgs boson, which acts as a mediator for the self-interaction between DM particles and antiparticles, as well as for the interaction between nucleons inside the NS. The presence of DM particles inside the NS and the interaction between DM and nucleons, mediated by Higgs, influences the EOS of NSs. It affects the stability and observable characteristics of NSs, such as their mass-radius relationship and tidal deformability \cite{10.1093/mnras/staa1435, Das_2021, 10.1093/mnras/stab2387}. Now, in the subsequent sections of this work, we systematically examine the dependence on $M_\chi$ in a range of $1 \le M_\chi\le 400$ GeV, along with a spectrum of Fermi momenta for the DM particles. Specifically, we explore the impact of certain ($M_{\chi}$, $k_f^{\text{DM}}$) range on the saturation parameters of nuclear matter, utilizing experimental observables. For NSs, we explore DM Fermi momenta range from 0.01 GeV to 0.15 GeV to derive constraints for ($M_{\chi}$, $k_f^{\text{DM}}$) plane based on astronomical observational data. These analyses are conducted using all six considered RMF parameter sets, although some of the parameter sets have already been excluded in light of the astronomical observations and terrestrial experiments, allowing us to thoroughly examine the parameter space and identify viable regions for DM within the context of NS observations.

\section{Results and discussion}
\label{sec:3}

\subsection{Parameter dependence of experimental observable}
\label{sec:3a}
\begin{figure}[tbp]
    \centering
    \includegraphics[width=\columnwidth]{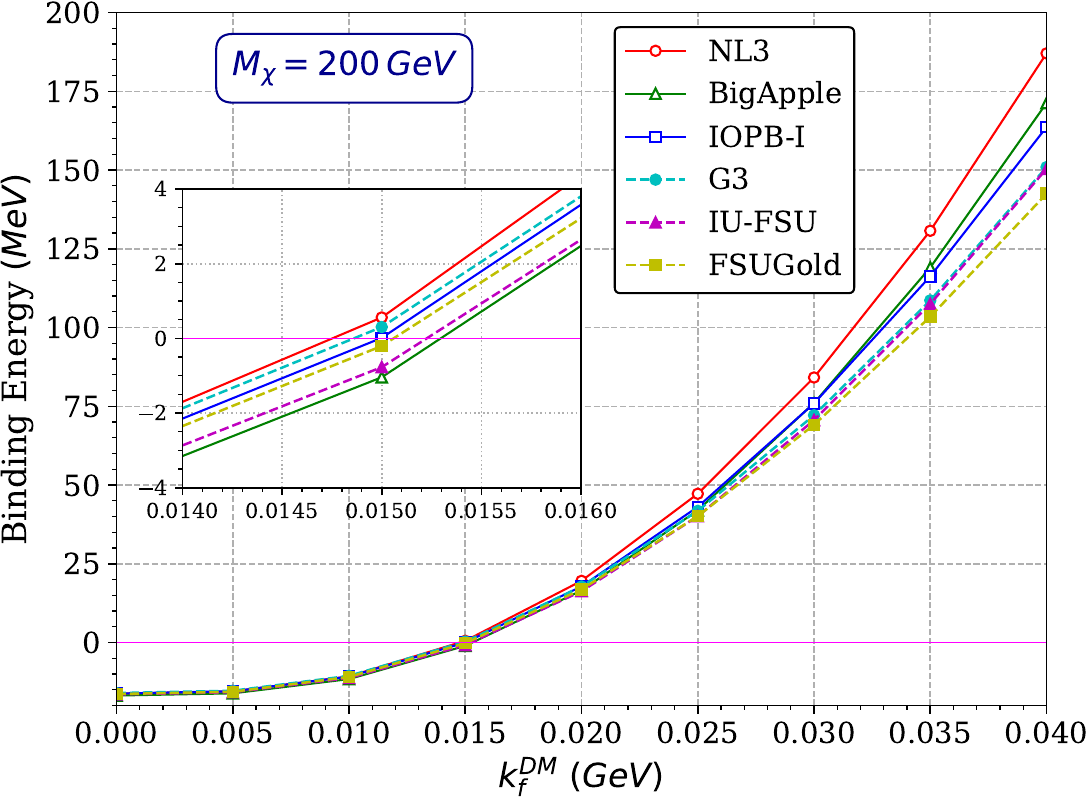} 
    \caption{For various EOSs, the binding energy of symmetric nuclear matter is shown as a function of $k_f^{\rm DM}$ with $M_{\chi}=200$ GeV. The inset is an enlarged view around the position where the binding energy becomes zero. }
    \label{fig:figure2}
\end{figure}

\begin{figure}[tbp]
    \centering
    \includegraphics[width=\columnwidth]{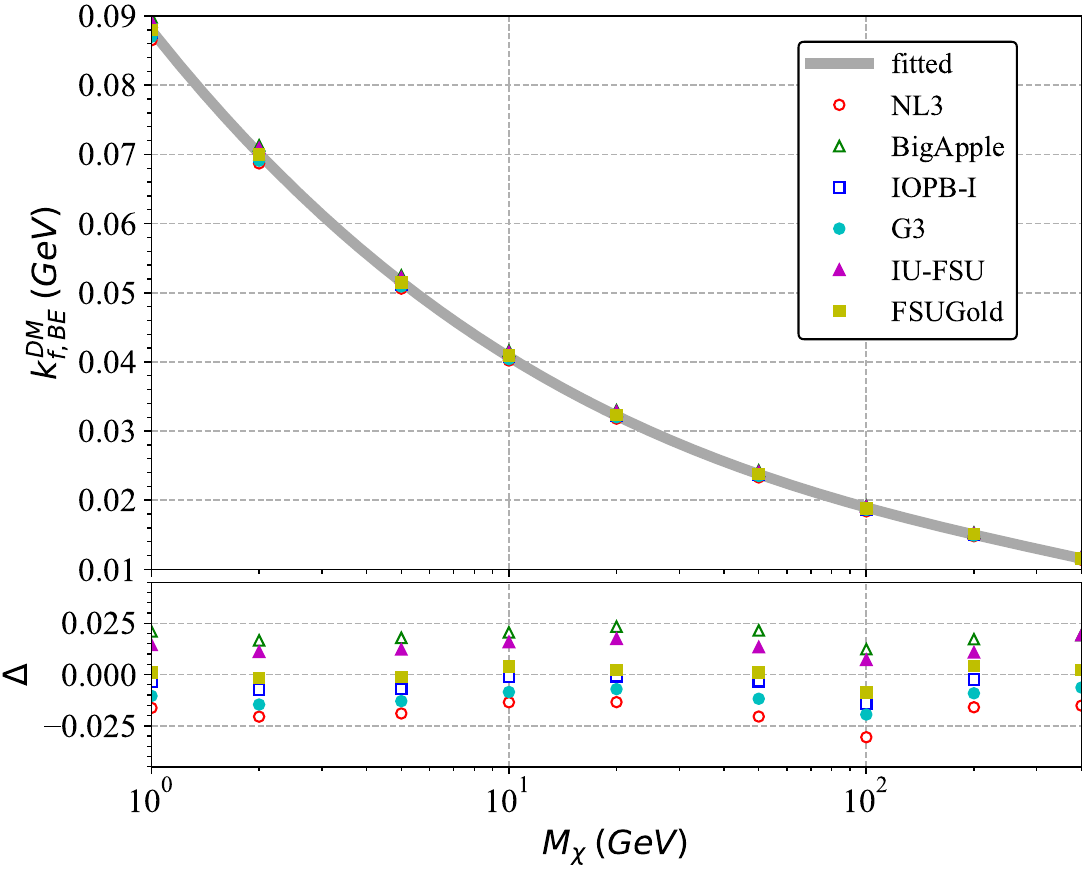} 
    \caption{In the top panel, the critical Fermi momentum of DM, $k_{f,{\rm BE}}^{\rm DM}$, is shown as a function of DM candidate mass for symmetric nuclear matter. $k_{f}^{\rm DM}$ should be satisfied by the condition of $k_f^{\rm DM}\le k_{f,{\rm BE}}^{\rm DM}$ so that the binding energy of symmetric nuclear matter at the saturation point becomes negative. The solid line denotes the fitting given by Eq.~(\ref{eq:k_fBE}). In the bottom panel, the relative deviation of the value of $k_{f,{\rm BE}}^{\rm DM}$ from the fitting for various EOS models.}
    \label{fig:figure3}
\end{figure}

In this section, we delve into how the inclusion of the DM model, as described earlier, influences the overall nuclear characteristics, basically, nuclear saturation properties, within the framework of the RMF theory. To begin with, we analyze the binding energy curve of symmetric nuclear matter as a function of the DM Fermi momentum for a fixed DM mass, i.e., $M_{\chi}$ = 200 GeV, across all the considered RMF parameter sets, with Fig. \ref{fig:figure2} illustrating this dependency. We observe that for a fixed DM mass, the nuclear saturation density ($n_{0}$) increases with increasing $k_{f}^{DM}$, shifting from the typical range of ($\sim 0.15 - 0.16$ ${\rm fm}^{-3}$) to higher values. Table \ref{table1} presents the saturation density of symmetric nuclear matter across different RMF parameter sets, illustrating the influence of DM characterized by various masses (0, 1, 10, 100 GeV) and Fermi momenta (0, 0.01, 0.02, 0.04 GeV). As seen, saturation density generally increases with both higher DM mass and Fermi momentum, indicative of the densification of nuclear matter in the presence of DM. This indicates that the presence of DM modifies the equilibrium properties of nuclear matter. Our observations reveal a critical phenomenon: for a fixed value of $M_{\chi}$, the binding energy of the DM-admixed symmetric nuclear matter transitions from negative to positive values as the DM momentum increases. This transition indicates that the nuclear matter becomes unbound beyond a certain threshold of DM Fermi momentum. The specific value of the DM Fermi momentum at which the binding energy reaches zero, termed the critical Fermi momentum for binding energy and represented by $k_{f,{\rm BE}}^{\rm DM}$, is crucial in understanding the stability of the nuclear matter in the presence of DM.
Furthermore, our analysis shows that this critical Fermi momentum ($k_{f,{\rm BE}}^{\rm DM}$) decreases as the DM particle mass increases. This inverse relationship highlights the significant impact of DM particle mass on the binding energy of nuclear matter. To provide a more quantitative understanding, the second plot (Fig. \ref{fig:figure3}) in this section presents the critical values of the DM Fermi momentum as a function of the $M_{\chi}$ for all RMF parameter sets. Remarkably, we find that the critical Fermi momentum ($k_{f,{\rm BE}}^{\rm DM}$) is independent of the RMF parameter sets, and follows a specific empirical relation given by:
\begin{equation} \label{eq:k_fBE}
    k_{f, {\rm BE}}^{\rm DM} = 0.08791 -0.06505x + 0.02048x^2 -0.002597 x^3,
\end{equation}
where $k_{f, {\rm BE}}^{\rm DM}$ is in the unit of GeV and $x$ is defined as $x\equiv \log_{10} [M_\chi /(1\ {\rm GeV})]$. The bottom panel of Fig. \ref{fig:figure3} shows the relative deviation, $\Delta$, of the actual values, $y_{\text{data}}$, from the value estimated with the empirical relation, $y_{\text{fit}}$, quantified by
\begin{equation}
  \Delta = (y_{\text{data}} - y_{\text{fit}})/y_{\text{data}}.  \label{eq:Delta}
\end{equation}
The maximum deviation from the empirical relation is found to be $\sim 2.5\,\%$ and that is also for a few cases, which is quite convincing and indicates that the empirical relation is a robust predictor of the critical Fermi momentum across different DM masses and RMF parameter sets.

In addition, we also explore how other nuclear saturation parameters are affected by the inclusion of the DM model within the RMF framework. 
\begin{table*}[]
\centering
\renewcommand{\tabcolsep}{0.355cm}
\renewcommand{\arraystretch}{1.75}
\caption{The saturation density, $n_0$, of symmetric nuclear matter, depending on the DM parameters, for various RMF parameter sets. For reference, the values without DM, i.e., $M_\chi=0$ and $k_f^{\rm DM}=0$, are also shown.}
\begin{tabular}{c|c|ccc|ccc|ccc}
\hline
\hline
\multirow{2}{*}{\makecell{DM Mass (GeV)  \\ \\ 
$k_{f}^{DM}$ (GeV)}} & 0 & \multicolumn{3}{c|}{$M_{\chi}=$ 1} & \multicolumn{3}{c|}{$M_{\chi}=$ 10} & \multicolumn{3}{c}{$M_{\chi}=$ 100} \\ \cline{1-11} 
 & 0 & \multicolumn{1}{c|}{0.01} & \multicolumn{1}{c|}{0.02} & 0.04 & \multicolumn{1}{c|}{0.01} & \multicolumn{1}{c|}{0.02} & 0.04 & \multicolumn{1}{c|}{0.01} & \multicolumn{1}{c|}{0.02} & 0.04 \\ \hline
NL3 & 0.148 & \multicolumn{1}{c|}{0.148} & \multicolumn{1}{c|}{0.149} & 0.156 & \multicolumn{1}{c|}{0.150} & \multicolumn{1}{c|}{0.158} & 0.196 & \multicolumn{1}{c|}{0.160} & \multicolumn{1}{c|}{0.203} & 0.295 \\ 
BigApple & 0.154 & \multicolumn{1}{c|}{0.155} & \multicolumn{1}{c|}{0.156} & 0.164 & \multicolumn{1}{c|}{0.156} & \multicolumn{1}{c|}{0.166} & 0.211 & \multicolumn{1}{c|}{0.169} & \multicolumn{1}{c|}{0.219} & 0.319 \\ 
IOPB-I & 0.149 & \multicolumn{1}{c|}{0.150} & \multicolumn{1}{c|}{0.151} & 0.160 & \multicolumn{1}{c|}{0.151} & \multicolumn{1}{c|}{0.162} & 0.207 & \multicolumn{1}{c|}{0.164} & \multicolumn{1}{c|}{0.215} & 0.339 \\
G3 & 0.148 & \multicolumn{1}{c|}{0.148} & \multicolumn{1}{c|}{0.149} & 0.157 & \multicolumn{1}{c|}{0.149} & \multicolumn{1}{c|}{0.159} & 0.207 & \multicolumn{1}{c|}{0.162} & \multicolumn{1}{c|}{0.217} & 0.374 \\ 
IU-FSU & 0.154 & \multicolumn{1}{c|}{0.154} & \multicolumn{1}{c|}{0.156} & 0.164 & \multicolumn{1}{c|}{0.156} & \multicolumn{1}{c|}{0.166} & 0.214 & \multicolumn{1}{c|}{0.169} & \multicolumn{1}{c|}{0.223} & 0.369 \\ 
FSUGold & 0.148 & \multicolumn{1}{c|}{0.148} & \multicolumn{1}{c|}{0.149} & 0.158 & \multicolumn{1}{c|}{0.149} & \multicolumn{1}{c|}{0.160} & 0.211 & \multicolumn{1}{c|}{0.163} & \multicolumn{1}{c|}{0.221} & 0.396 \\ 
\hline
\hline
\end{tabular}
\label{table1}
\end{table*}
For the EOS derived from any phenomenological model, the bulk energy per baryon, $\mathcal{E}/n_{B}$, for zero-temperature nuclear matter can be expressed as an expansion around the saturation density of symmetric nuclear matter. This expansion is formulated in terms of the baryon number density, $n_{B}$, and the neutron-proton asymmetry, $\alpha$, defined as $\alpha = (n_{n} - n_{p})/n_{B}$, with $n_{n}$ and $n_{p}$ representing the neutron and proton number densities, respectively. The expression is given by \cite{annurev:/content/journals/10.1146/annurev.ns.31.120181.002005}:

\begin{figure*}[tbp]
    \centering
    \includegraphics[width=\textwidth]{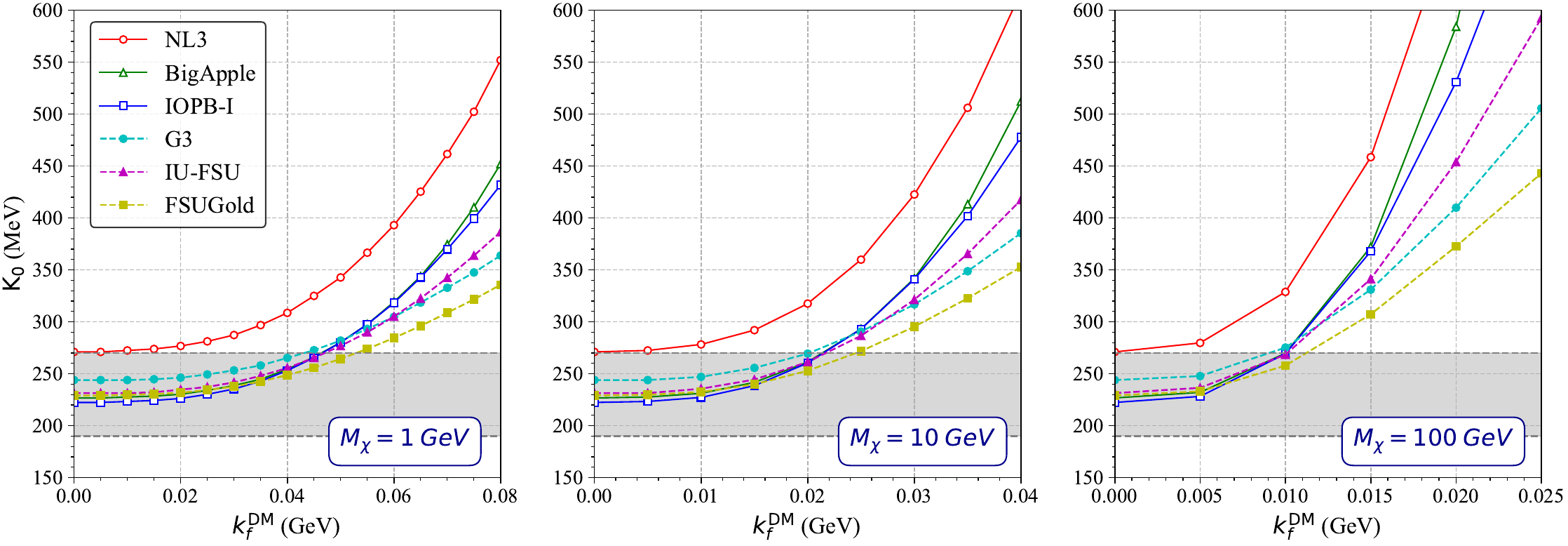} 
    \caption{$k_f^{\rm DM}$ dependence of $K_0$ for various RMF parameter sets. The panels from left to right correspond to the results for $M_\chi=1$, 10, and 100 GeV. The shaded region denotes the constraint on $K_0$ obtained experimentally, i.e., $K_0=230 \pm 40$ MeV \cite{PhysRevC.88.034319}. It is observed that the range of critical $k_f^{\rm DM}$ varies with $M_\chi$.}
    \label{fig:figure4}
\end{figure*}

\begin{align}
    \frac{\mathcal{E}}{n_{B}} =& \,w_{0} + \frac{1}{2} K_0 u^2 \nonumber \\
       &+ \left( J_{0} + L_{\rm sym,0} u +{\cal O}(u^2) \right)\alpha^2 + \mathcal{O} \left(u^3, \alpha^4 \right),  \label{eq:bulk}  
\end{align}
where $u$ is defined as $u\equiv (n_B - n_0)/(3\,n_0)$, $K_{0}$ is incompressibility, $J_{0}$ is the symmetry energy, and $L_{\rm sym,0}$ is known as the slope parameter at saturation density.

The incompressibility is crucial as it indicates how resistant the nuclear matter is to compression at saturation density; a higher $K_0$ implies a stiffer EOS. Figure \ref{fig:figure4} illustrates $K_0$ as a function of $k_f^{\text{DM}}$ for different DM masses: 1 GeV, 10 GeV, and 100 GeV from the left to right; alongside empirical values with the shaded region adopted from Ref.~\cite{PhysRevC.88.034319}, i.e., $K_0=230\pm 40$ MeV, even though more sever constraint on $K_0$ has also been reported~\cite{Shlomo2006}. Our observations reveal that for a fixed DM mass, the incompressibility coefficient, $K_0$, increases with increasing $k_f^{\text{DM}}$ across all RMF parameter sets. This trend suggests that the EOS becomes stiffer as the DM Fermi momentum rises. Additionally, for a fixed $k_f^{\text{DM}}$, an increase in $K_0$ is observed with increasing $M_{\chi}$, which indicates that the presence of more massive DM particles also contributes to a stiffer EOS. However, this apparent increase in stiffness with higher DM mass or Fermi momentum presents a paradox when considering NS properties. Typically, a stiffer EOS would lead to a higher maximum mass for NSs. Contrary to this expectation, our results, as we will discuss in the next section, show that the maximum mass of NS decreases with increasing DM mass or Fermi momentum. This discrepancy suggests that the interpretation of $K_0$ as an indicator of stiffness may be more complex in the context of DM-admixed nuclear matter. One possible explanation for this apparent contradiction lies in the change in saturation density with the inclusion of DM. As shown in Table \ref{table1}, the saturation density increases with increasing DM Fermi momentum for a fixed value of $M_{\chi}$. Consequently, the value of $K_0$ obtained at this elevated saturation density is higher compared to the case without DM inclusion. This increase in $K_{0}$ at a higher saturation density does not necessarily indicate a stiffer EOS in the traditional sense, but rather reflects the altered equilibrium properties of the nuclear matter in the presence of DM.

\begin{figure*}[tbp]
    \centering
    \includegraphics[width=\textwidth]{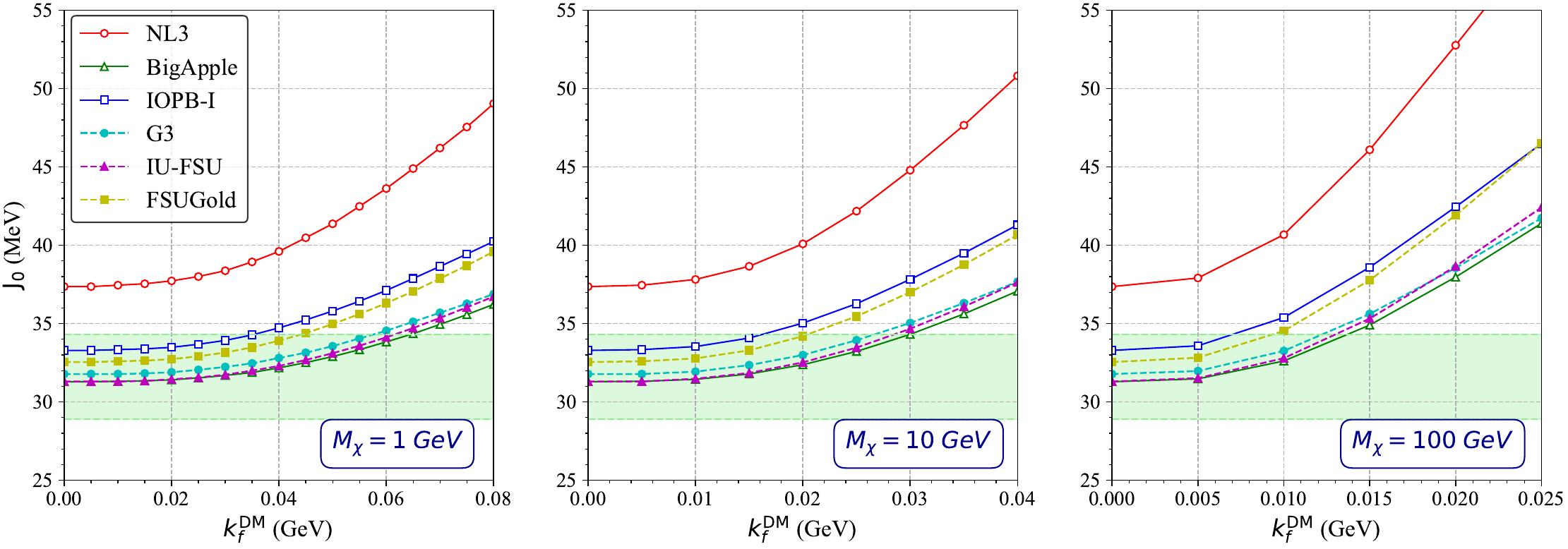} 
    \caption{$k_f^{\rm DM}$ dependence of $J_0$ for various RMF parameter sets. The panels from left to right correspond to the results with $M_\chi=1$, 10, and 100 GeV. The shaded region denotes the constraint on $J_0$ obtained experimentally, i.e., $J_0=31.6 \pm 2.7$ MeV \cite{Li:2019xxz}. We note that the RMF model with NL3 parameter set is excluded even without the DM contribution, i.e., $k_f^{\rm DM}=0$.}
    \label{fig:figure5}
\end{figure*}

\begin{figure*}[tbp]
    \centering
    \includegraphics[width=\textwidth]{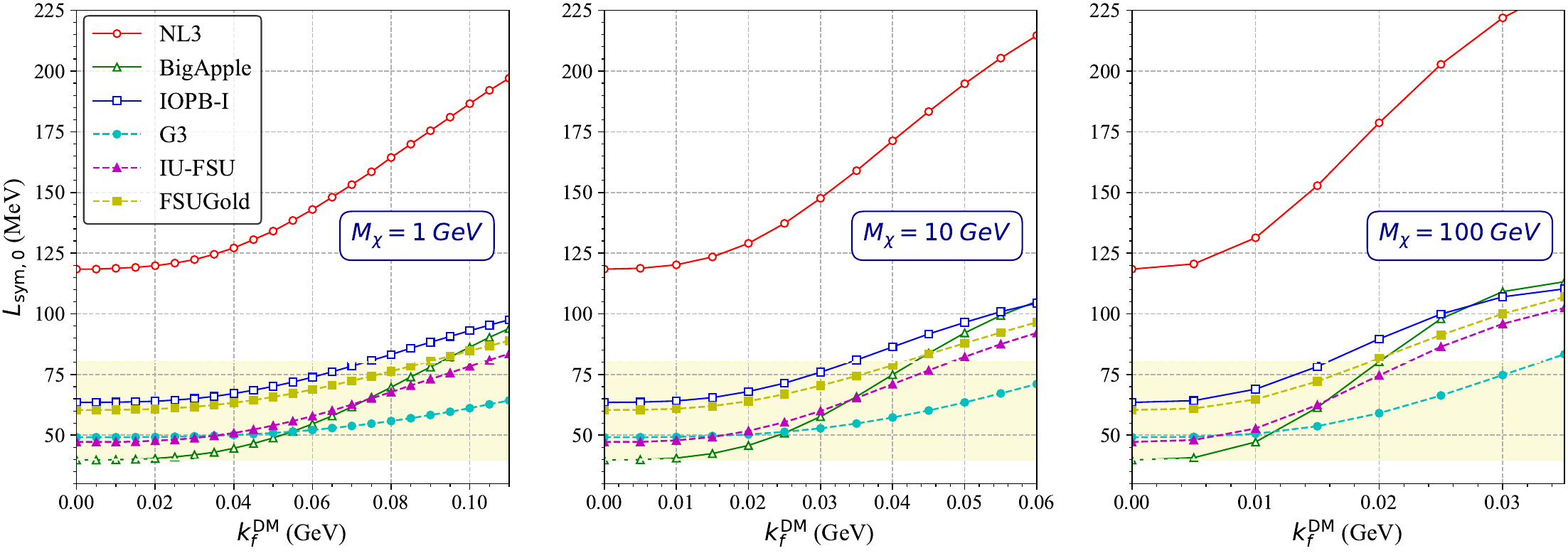} 
    \caption{$k_f^{\rm DM}$ dependence of $L_{\rm sym,0}$ for various RMF parameter sets. The panels from left to right correspond to the results with $M_\chi=1$, 10, and 100 GeV. The shaded region denotes the constraint on 
    $L_{\rm sym,0}$ obtained experimentally, i.e., $L_{\rm sym,0}=60 \pm 20$ MeV \cite{Li:2019xxz}. We note that the RMF model with NL3 parameter set is excluded even without the DM, i.e., $k_f^{\rm DM}=0$.}
    \label{fig:figure6}
\end{figure*}

The subsequent series of plots (Figs.~\ref{fig:figure5} and \ref{fig:figure6}) examine two critical parameters at saturation density of symmetric nuclear matter: the symmetry energy ($J_0$) and the slope parameter ($L_{\text{sym},0}$). Both parameters are analyzed across different DM masses (1, 10, and 100 GeV) for all adopted RMF parameter sets. The shaded regions in these plots represent the experimentally obtained values, specifically $J_0 = 31.6 \pm 2.7$ MeV for the symmetry energy and $L_{\text{sym},0} = 60 \pm 20$ MeV for the slope parameter \cite{Li:2019xxz}\footnote{There is still a large uncertainty in the experimental constraints on $L_{\text{sym},0}$, although we refer its fiducial value here. For example, it has been reported $42 \le L_{\text{sym},0} \le 117$ MeV \cite{PhysRevLett.126.162701} or $L_{\text{sym},0}=106\pm37$ MeV \cite{PhysRevLett.126.172503}.}.

Starting with the symmetry energy ($J_0$), one notable observation is that the NL3 parameter set predicts $J_0$ value outside the experimentally determined range even in the absence of any DM component (i.e., with $k_f^{\text{DM}}$ = 0). This deviation highlights a potential limitation or peculiarity of the NL3 parameter set in accurately modeling the symmetry energy of nuclear matter. Another significant observation from our calculations is the trend of rising $J_0$ values with $k_f^{\text{DM}}$ and DM mass. For instance, for $M_{\chi}$ = 10 GeV, the $J_0$ value at a DM Fermi momentum of 0.15 GeV falls within the experimentally determined range. However, for the same DM Fermi momentum but with a DM particle mass of 100 GeV, the $J_0$ value is considerably higher and deviates from the experimental range. This trend is consistent across all RMF parameter sets, indicating that higher DM masses tend to push the symmetry energy beyond the experimentally established bounds. Turning to the $L_{\text{sym},0}$, we observe similar trends. The NL3 parameter set again exhibits values outside the experimentally determined range even without any DM component. As with $J_0$, the $L_{\text{sym},0}$ values increase with rising DM Fermi momentum for a fixed DM mass. 

These observations suggest that the presence of DM can significantly alter both the symmetry energy and the slope parameter at saturation density. The increase in $J_0$ and $L_{\text{sym},0}$ with DM Fermi momentum and mass imply that the nuclear matter becomes more asymmetric in the sub-nuclear density and the density dependence of the symmetry energy becomes steeper. This could have profound implications for the structure and properties of NSs and other astrophysical objects. The deviation of these parameters from the experimental range, particularly at higher DM masses, underscores the importance of considering DM effects in nuclear matter studies. Currently, nuclear experiments do not account for the potential influence of DM on nuclear properties. The findings from our study suggest that future experimental and theoretical investigations should incorporate DM effects to provide a more comprehensive understanding of nuclear matter under the presence of DM. While the consideration of DM in nuclear experiments may seem hypothetical at this stage, the significant deviations observed in our study indicate that DM could play a crucial role in shaping the properties of nuclear matter.

\subsection{Constraints from astronomical observations}
\label{sec:3b}
This section elucidates the constraints on DM mass and momentum space derived from astronomical data on the NS mass and radius. Utilizing a series of detailed plots, we explore the methodological steps to establish these constraints, incorporating RMF parameter sets in line with recent astrophysical measurements. Guided by critical observational benchmarks — such as the maximum mass range for PSR J0740+6620, the NICER-determined mass-radius range with 95\% credibility for PSR J0030+0451, and tidal deformability metrics for a canonical NS (1.4 $M_{\odot}$) — our analysis provides a quantitative framework to assess the DM space within NSs. This approach not only ensures rigorous derivation of DM constraints but also integrates these findings effectively with existing observational data, enhancing our understanding of how DM properties influence NS observations.

\begin{figure*}[tbp]
    \centering
    \includegraphics[width=\textwidth]{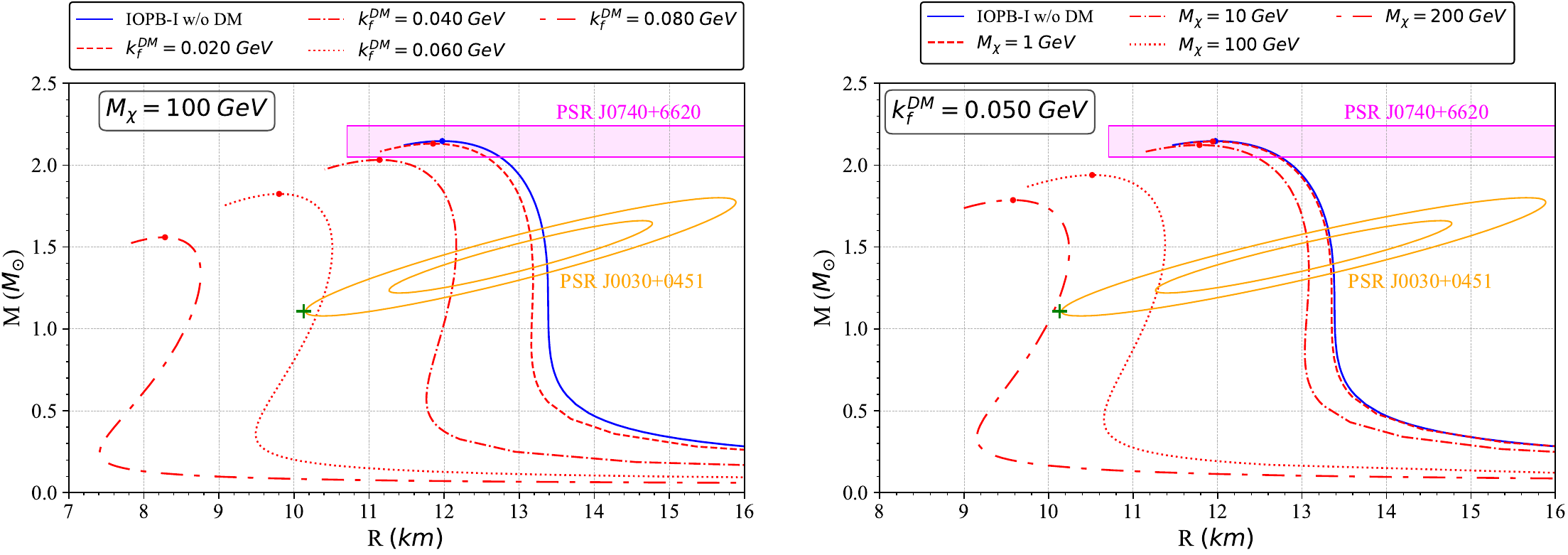} 
    \caption{The mass and radius of DM admixed NS models for various DM parameters, adopting the IOPB-I parameter set. The left panel is the $k_f^{\rm DM}$ dependence with $M_\chi=100$ GeV, while the right panel is the $M_\chi$ dependence with $k_f^{\rm DM}=0.05$ GeV. As in Fig.~\ref{fig:figure1}, the mass of PSR J0740+6620 and the constraints on the NS mass and radius for PSR J0030+0451 obtained from the NICER observations are shown. The plus denotes the NS model with $M=1.108M_\odot$ and $R=10.13$ km, which corresponds to the leftmost boundary in the 95\%  credibility for PSR J0030+0451.}
    \label{fig:figure7}
\end{figure*}

Exploring the impact of DM on the mass-radius ($M$-$R$) relationships of NSs, detailed calculations reveal how changes in DM parameters can significantly alter NS properties. One initial analysis focuses on the effects of varying DM Fermi momenta ($k_f^{\text{DM}}$) with a fixed DM mass of 100 GeV, alongside a reference curve for the IOPB-I parameter set without DM (the left panel of Fig.~\ref{fig:figure7}). It shows that increasing $k_f^{\text{DM}}$, from 0.020 GeV to 0.080 GeV, leads to a noticeable contraction in the radius for each mass of NS, suggesting denser configurations. This trend is accompanied by a slight reduction in the maximum mass of NSs, indicating a softening of the EOS due to the presence of DM.
Another set shown in the right panel of Fig.~\ref{fig:figure7} examines the role of DM mass on NS properties, holding $k_f^{\rm DM}$ constant at 0.050 GeV and varying the DM particle mass from 1 GeV to 200 GeV. These results show that increasing the DM mass leads to more substantial reductions in the radius of NSs at similar mass points, especially noticeable for higher DM masses such as 100 GeV and 200 GeV. This pronounced effect suggests that heavier DM particles contribute to a significantly denser NS core.

In NS models that include DM, the mass of DM particles plays a crucial role in defining the gravitational binding and stability of the star. Heavier DM particles, such as those in the range of 100 GeV to 200 GeV, introduce significant additional gravitational mass within the core of NSs. This additional mass contributes to the overall gravitational pull exerted by the star, which can lead to more compact and denser star configurations as observed in the $M$-$R$ profiles. This increase in central mass also affects the NS’s EOS. Typically, a stiffer EOS, which supports higher maximum mass against gravitational collapse, might be counteracted by the inclusion of heavy DM particles, softening the EOS due to the additional gravitational effects and altered nuclear interactions influenced by the dense DM core. Despite the increased density and gravitational binding provided by heavier DM, this might paradoxically result in a lower maximum mass for the NS as the increased DM mass begins to dominate the dynamics, potentially hastening gravitational collapse under certain conditions. Understanding these dynamics is crucial, especially in light of observational data from GW events and NS measurements, which can provide empirical constraints on the theories regarding the interaction of DM with ordinary matter in extreme astrophysical environments.
\begin{figure}[tbp]
    \centering
    \includegraphics[width=\columnwidth]{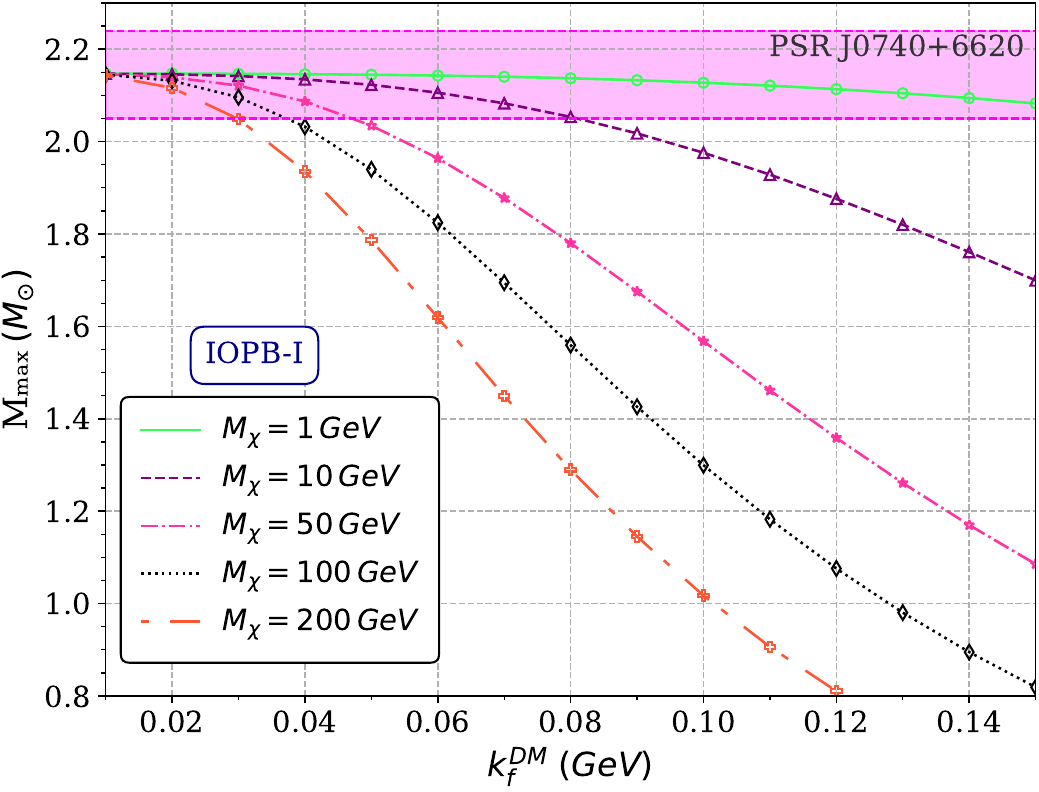} 
    \caption{Dependence of the maximum mass of NSs on the DM parameter, $k_f^{\rm DM}$ and $M_\chi$, using the IOPB-I parameter set. For reference, the mass of PSR J0740+6620 is also shown. }
    \label{fig:figure8}
\end{figure}

The next plot (Fig.~\ref{fig:figure8}) reveals the relationship between the maximum mass of NSs and varying levels of DM Fermi momentum, for several fixed DM particle masses. The data show a range of DM masses from 1 GeV to 200 GeV considering the IOPB-I RMF parameter set. A consistent trend observed across all curves is that the maximum mass of NS decreases as $k_f^{\rm DM}$ increases. This behavior is reflective of the softening of the EOS when higher DM content is factored into the NS model. Furthermore, the maximum NS mass for each DM mass scenario is compared against the mass limits of PSR J0740+6620, which has a well-established observational mass constraint. For each DM mass profile, a critical value of $k_f^{\rm DM}$ is extracted at which the maximum mass of the NS drops below the observational lower mass limit of PSR J0740+6620. These critical Fermi momentum values represent thresholds beyond which the DM parameters would lead to NS configurations that are not supported by current observational data.
\begin{figure}[tbp]
    \centering
    \includegraphics[width=\columnwidth]{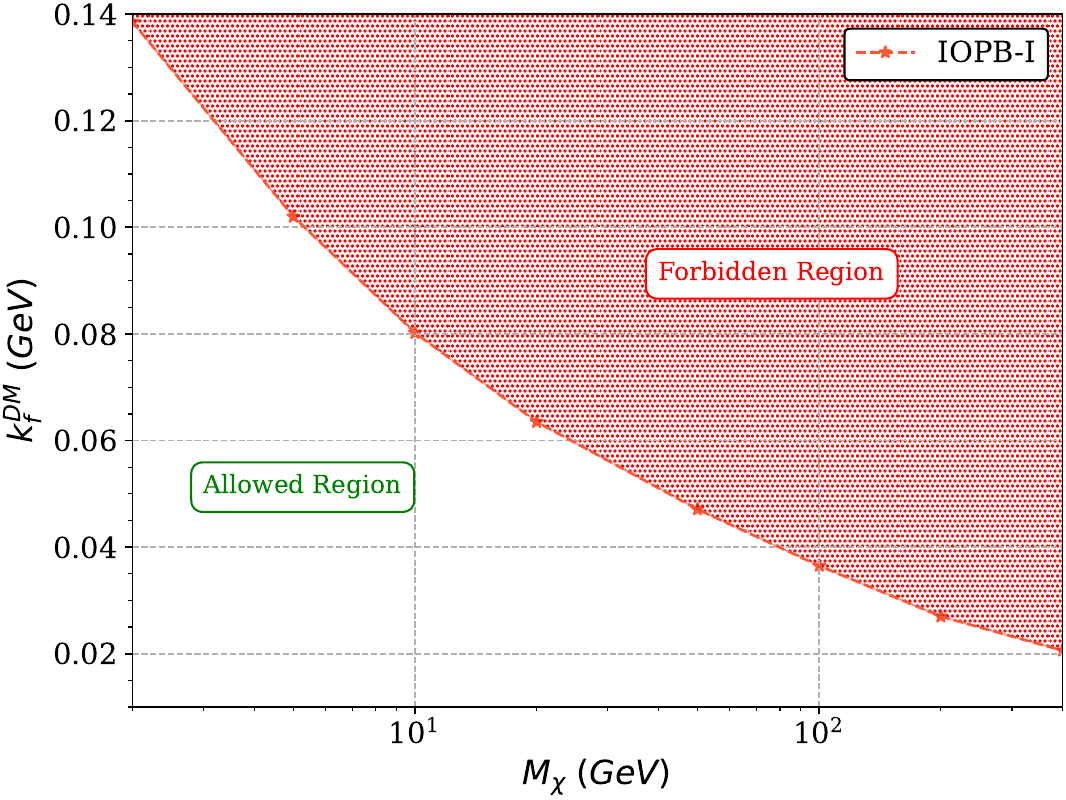} 
    \caption{Allowed parameter space in the $(M_\chi,k_f^{\rm DM})$-plane obtained from mass limits of PSR J0740+6620 data shown in Fig.~\ref{fig:figure8}, using the IOPB-I parameter set.}
    \label{fig:figure9}
\end{figure}
In the subsequent plot, these critical $k_f^{\rm DM}$ values are plotted as a function of $M_{\chi}$. This figure (Fig. \ref{fig:figure9}) effectively divides the DM parameter space into ``forbidden" and ``allowed" regions.
\begin{figure*}[tbp]
    \centering
    \includegraphics[width=\textwidth]{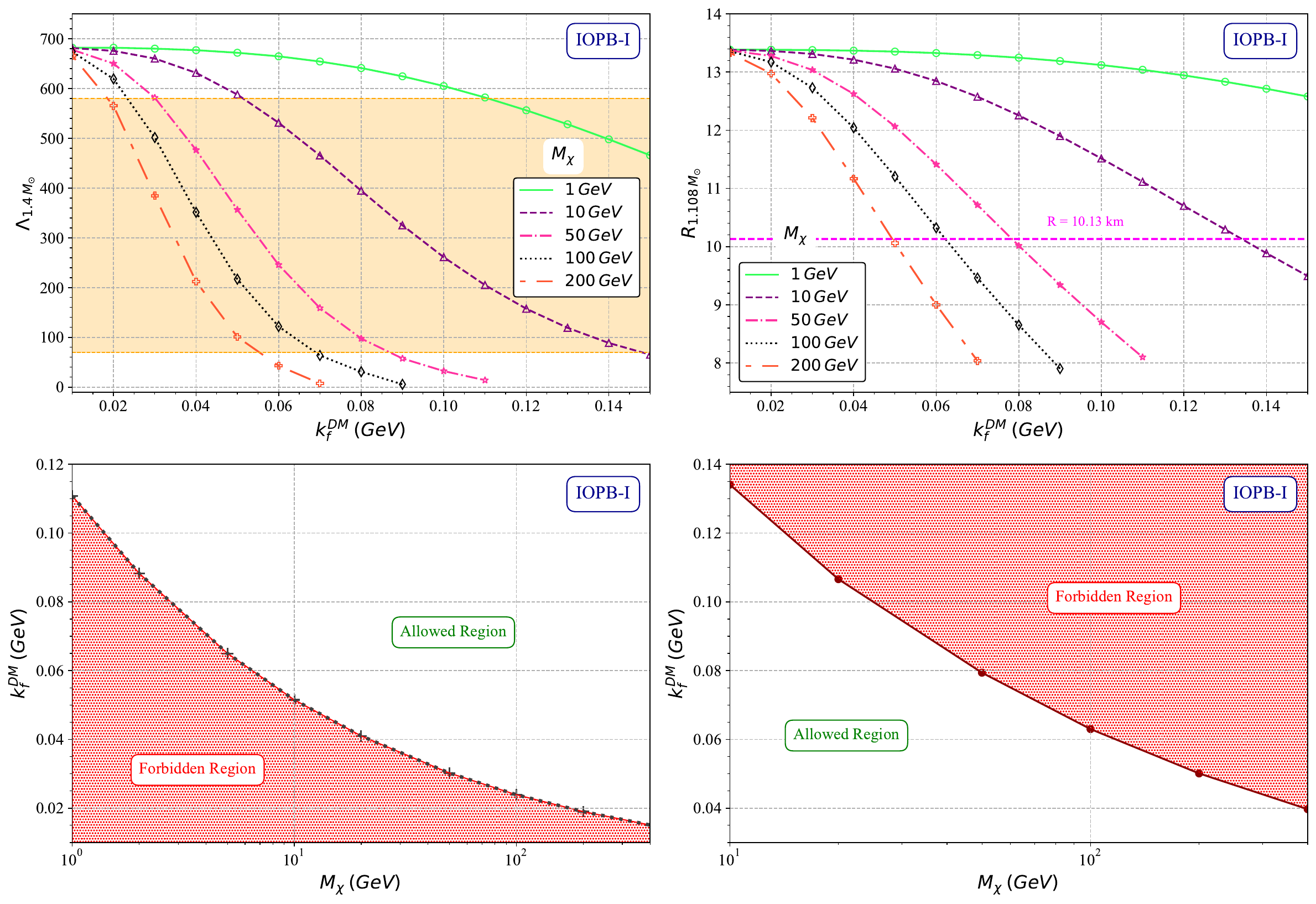} 
    \caption{The left-top and right-top panels are respectively the dependence of the dimensionless tidal deformability for the $1.4M_\odot$ NS model, and the NS radius for $1.108M_\odot$ with various DM parameters, adopting the IOPB-I parameter set. The shaded region in the left-top panel denotes the constraint on $\Lambda_{1.4M_\odot}$ obtained from the GW170817, i.e., $\Lambda_{1.4M_\odot} = 190^{+390}_{-120}$ \cite{PhysRevLett.121.161101}, while the horizontal dashed line in the right-top panel denotes the minimum radius for $1.108M_\odot$ NS constrained from the NICER for PSR J0030+0451  (also see Fig.~\ref{fig:figure7}). Considering these constraints, the value of $\Lambda_{1.4M_\odot}$ should be at least less than $580$, which corresponds to the upper boundary of the shaded region in the left-top panel and $R_{1.108M_\odot}$ should be larger than 10.13 km, which give us the allowed region in the ($M_\chi$, $k_f^{\rm DM}$) parameter space as shown in the left-bottom and right-bottom panels.}
    \label{fig:figure10}
\end{figure*}
The forbidden region encompasses DM parameter combinations where the Fermi momentum exceeds the critical value for a given DM mass, thus yielding NS maximum masses that fall below the empirical mass of PSR J0740+6620. Conversely, the allowed region contains the combinations of DM mass and Fermi momentum that comply with the observed NS mass limits. This analysis imposes tight constraints on the permissible DM properties within NSs specifically regarding their maximum mass by integrating observational data from PSR J0740+6620 with the adopted RMF theoretical model.

Our further study investigates the impact of DM on the tidal deformability of canonical NSs ($\Lambda_{1.4 M_{\odot}}$). The left-top panel of Fig. \ref{fig:figure10} showcases the dependence of $\Lambda_{1.4 M_{\odot}}$ on DM Fermi momentum for various DM masses ranging from 1 GeV to 200 GeV with the IOPB-I parameter set. As $k_f^{\rm DM}$ increases, $\Lambda_{1.4 M_{\odot}}$ consistently decreases, for each case of fixed $M_{\chi}$. This trend is highlighted against the observational constraints from GW170817, which provide a shaded region representing the empirically determined limits for $\Lambda_{1.4 M_{\odot}}$, i.e., $= 190^{+390}_{-120}$ \cite{PhysRevLett.121.161101}. From this data, we extract critical $k_f^{\rm DM}$ values at which the tidal deformability reaches the upper limit of the observed constraint, marking these as thresholds beyond which the presence of DM renders the NS insufficiently deformable according to current observational data. Plotting these critical $k_f^{\rm DM}$ values against $M_{\chi}$ (left-bottom panel of Fig. \ref{fig:figure10}), we again delineate forbidden and allowed regions in the parameter space of DM in light of canonical tidal deformability. 
\begin{figure}[tbp]
    \centering
    \includegraphics[width=\columnwidth]{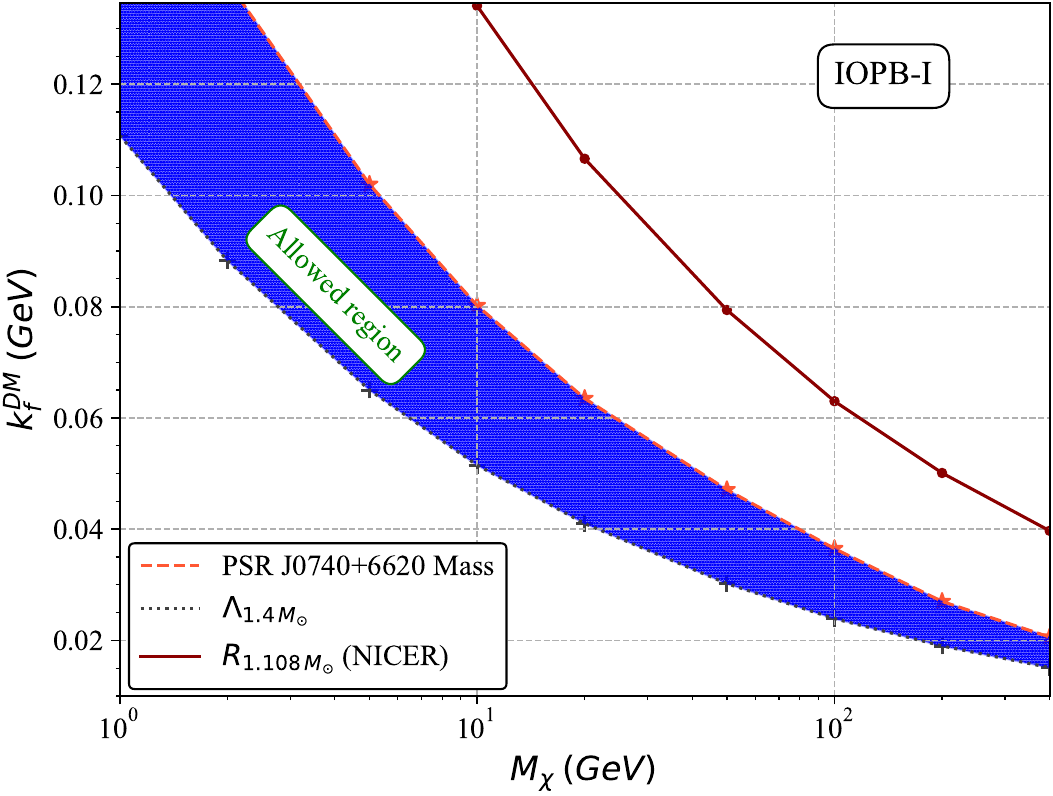} 
    \caption{The allowed region in the ($M_\chi$, $k_f^{\rm DM}$) parameter space, obtained from the mass of PSR J0740+6620, the dimensionless tidal deformability constrained from GW170817, and the NS mass and radius constraint obtained from NICER observation for PSR J0030+0451, adopting the IOPB-I parameter set, derived from the overlapped space among the three allowed regions shown in Fig.~\ref{fig:figure9} and the left-bottom and right-bottom panels in Fig.~\ref{fig:figure10}.}
    \label{fig:figure11}
\end{figure}
The forbidden region includes parameter combinations where $k_f^{\rm DM}$ falls below the critical value for the corresponding value of $M_{\chi}$, resulting in a tidal deformability that does not comply with empirical observations. Conversely, the allowed region encompasses parameter combinations above these critical values, adhering to the observational limits and suggesting feasible scenarios for DM characteristics within NSs.

Further refining our analysis, we examine the radius of a NS with a mass of 1.108 $M_{\odot}$ across different DM Fermi momenta and masses.
\begin{figure*}[tbp]
    \centering
    \includegraphics[width=\textwidth]{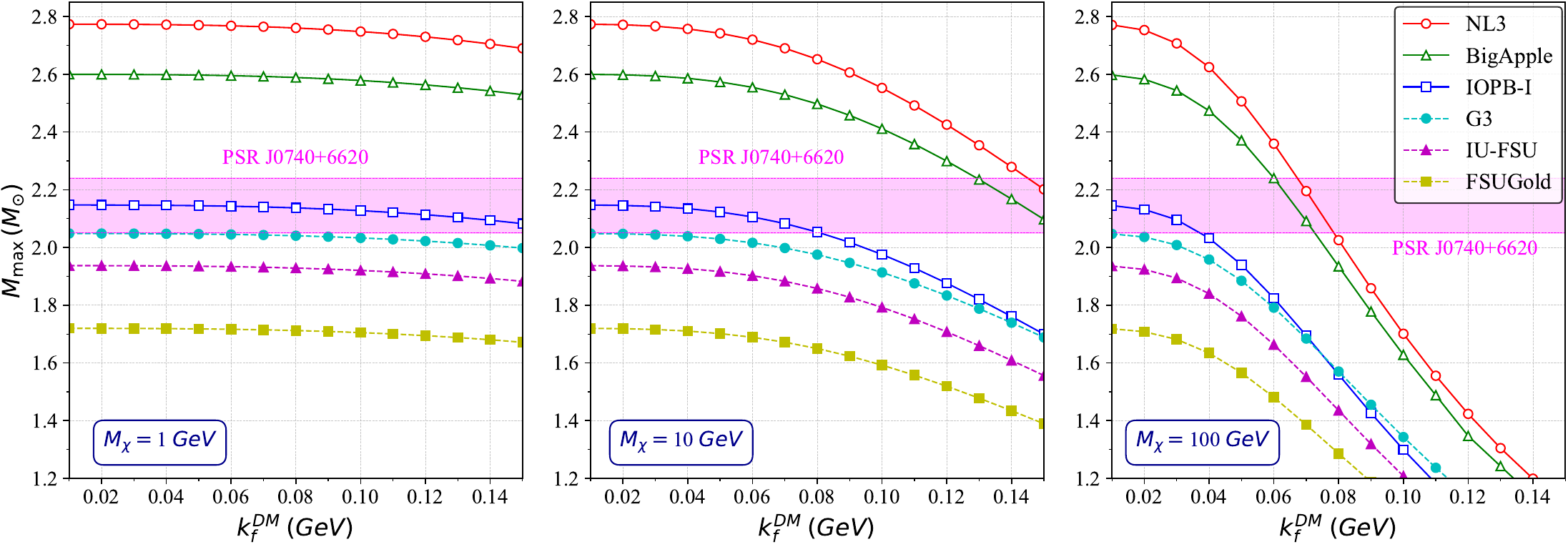} 
    \caption{Same as in Fig.~\ref{fig:figure8}, but for various RMF parameter sets.}
    \label{fig:figure12}
\end{figure*}
This investigation aligns with recent NICER observations of PSR J0030+0451, which dictate that the radius of such a NS must exceed 10.13 km. Our findings indicate that as $k_f^{\rm DM}$ and $M_{\chi}$ increase, the radius of the NS decreases, shown in the right-top panel of Fig.~\ref{fig:figure10}, potentially violating the NICER constraint. This relationship is crucial for understanding the structural integrity of NSs under the influence of DM. 
\begin{figure*}[tbp]
    \centering
    \includegraphics[width=\textwidth]{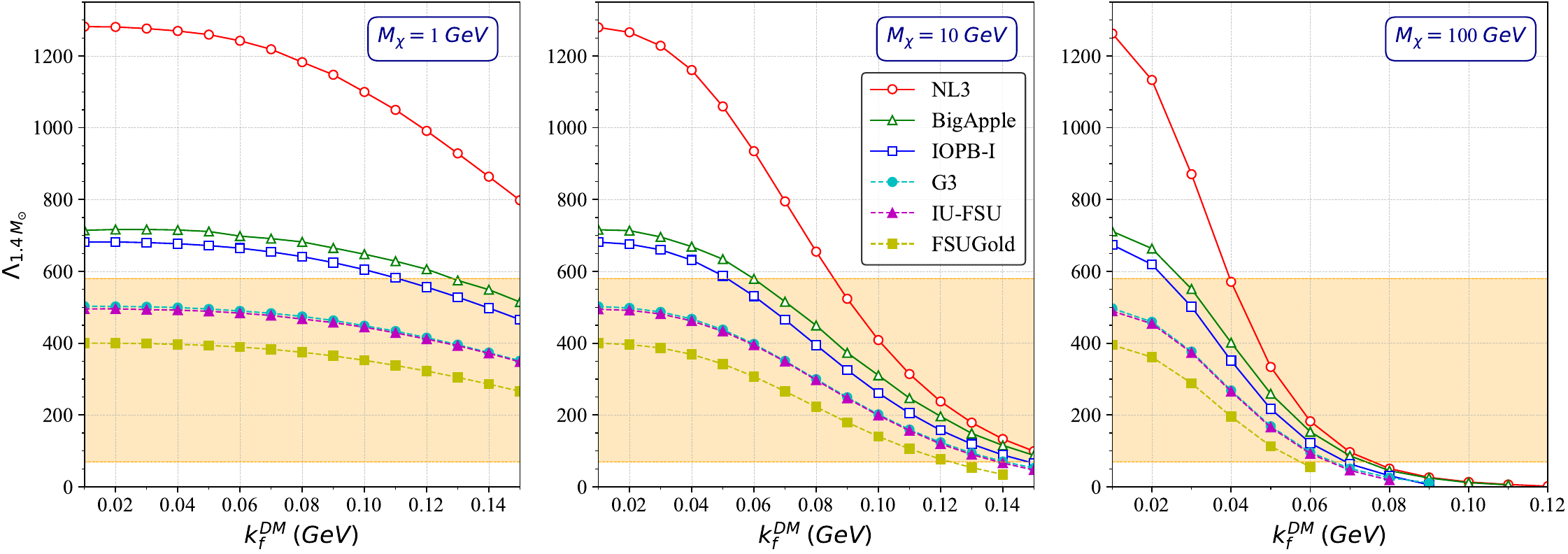} 
    \caption{Same as the left-top panel in Fig.~\ref{fig:figure10}, but for various RMF parameter sets.}
    \label{fig:figure13}
\end{figure*}
The critical $k_f^{\rm DM}$ values ensuring compliance with the NICER radius constraint are plotted against $M_{\chi}$ (right-bottom panel of Fig.~\ref{fig:figure10}), establishing a new set of forbidden and allowed regions. The forbidden region in this context includes DM parameter combinations leading to a NS radius smaller than the NICER threshold, while the allowed region includes those that maintain or exceed this limit.

In the comprehensive analysis, we consolidate findings from three critical observational constraints — maximum mass of PSR J0740+6620, tidal deformability of 1.4 $M_{\odot}$ star constrained from the GW170817, and NICER's constraints on the radius of a 1.108 $M_{\odot}$ NS — to delineate the permissible $k_f^{\rm DM}$  as a function of $M_{\chi}$ for the IOPB-I parameter set. Each constraint independently imposes limits on $k_f^{\rm DM}$ for varying DM masses, which are plotted together in Fig. \ref{fig:figure11} to identify a common allowed region for DM characteristics that are consistent with all observed astrophysical phenomena.
\begin{figure*}[tbp]
    \centering
    \includegraphics[width=\textwidth]{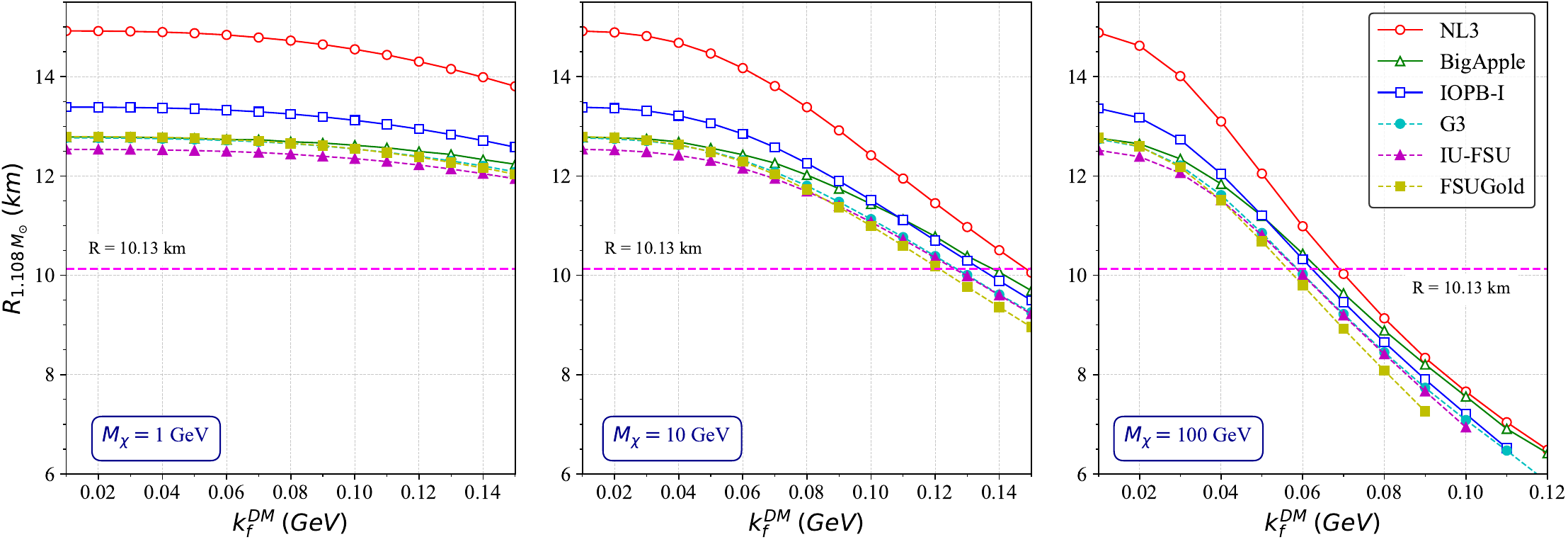} 
    \caption{Same as the right-top panel in Fig.~\ref{fig:figure10}, but for various RMF parameter sets.}
    \label{fig:figure14}
\end{figure*}
The intersection of the allowed regions from these three constraints establishes the final permissible range for $k_f^{\rm DM}$ across a wide range of $M_{\chi}$. Figure \ref{fig:figure11} clearly illustrates that the most restrictive constraints are typically provided by the tidal deformability and PSR J0740+6620 mass limit, which limit the higher DM Fermi momenta more stringently than the NICER radius observations alone. The final allowed region, therefore, represents a conservative and robust set of DM parameters that simultaneously satisfies all three observational constraints within the framework of the IOPB-I parameter set.

In addition to the parameters in the DM model considered in this study, such as $k_f^{\rm DM}$ and $M_\chi$, the EOS is another important input physics for considering the NS mass and radius, as shown in Fig.~\ref{fig:figure1}. That is, in Fig.~\ref{fig:figure11} we derive the $(k_f^{\rm DM}, M_\chi)$ parameter space, which can be excluded from the astronomical constraints, but this parameter space should depend on the adopted EOS as well. Now, in further discussion, we will examine how the allowed region in the $(k_f^{\rm DM}, M_\chi)$ parameter space depends on the EOS. The plots across Figs.~\ref{fig:figure12}, \ref{fig:figure13}, and \ref{fig:figure14} provide a rich dataset to examine how the inclusion of DM influences NS properties such as maximum mass, tidal deformability, and radius across adopted RMF parameter sets (NL3, BigApple, IOPB-I, G3, IU-FSU, and FSUGold) in this study. 

Figure \ref{fig:figure12} intricately illustrates how the maximum mass of NSs varies as a function of DM Fermi momentum for three distinct DM masses, i.e., 1, 10, and 100 GeV. Stiffer RMF models like NL3 and BigApple support maximum masses particularly at lower $k_f^{\rm DM}$ values with higher $M_{\chi}$, which align with the observational mass limits of PSR J0740+6620. This suggests that stiffer EOS can support larger NS masses, even under significant DM influences, particularly for higher DM masses which inherently present more gravitational challenges. Conversely, softer RMF models such as FSUGold and IU-FSU show inherent limitations, with their NS maximum mass predictions falling below the PSR J0740+6620 limit even in the absence of DM.
\begin{figure*}[tbp]
    \centering
    \includegraphics[width=\textwidth]{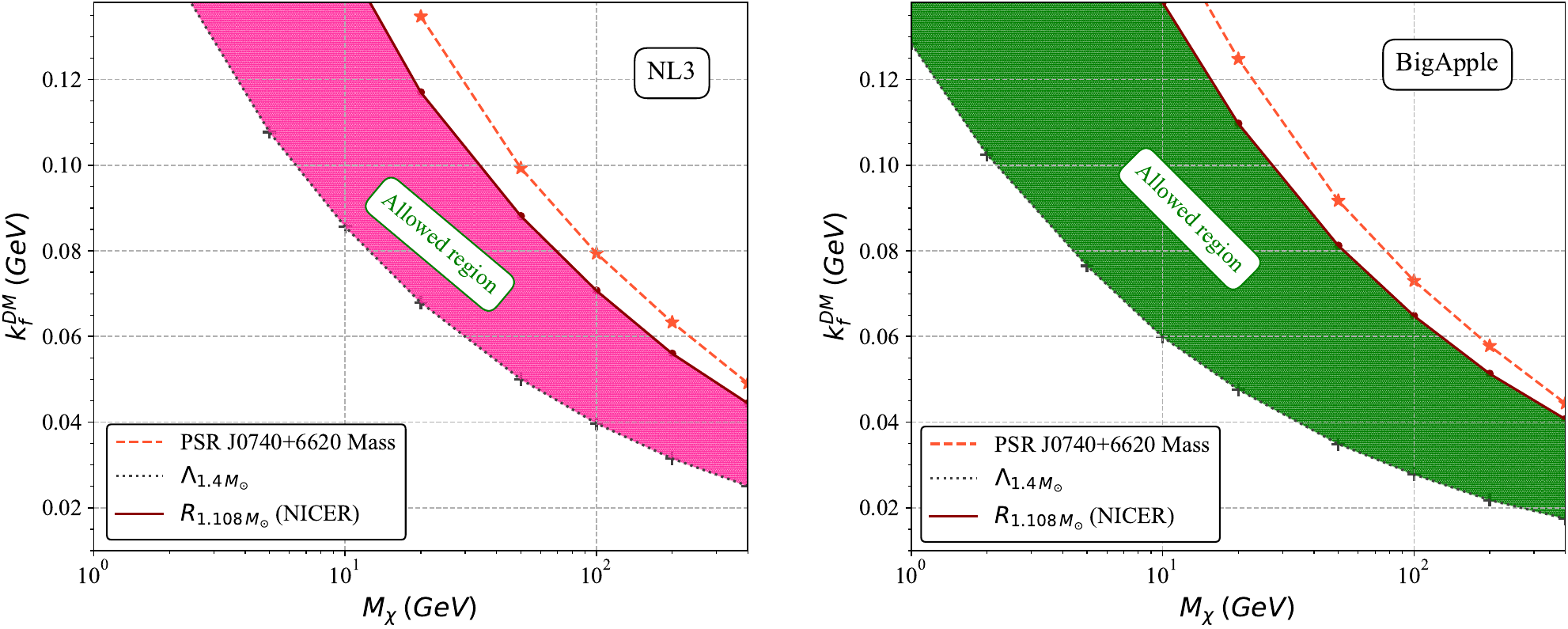} 
    \caption{Same as Fig.~\ref{fig:figure11}, but for the NL3 (the BigApple) parameter set in the left (right) panel. We note that the right boundary in Fig.~\ref{fig:figure11} for the IOPB-I parameter set is given by the constraint on the mass of PSR J0740+6620, while those for the NL3 and BigApple parameter sets are given by the constraint from the NICER observation for PSR J0030+0451.}
    \label{fig:figure15}
\end{figure*}
The inclusion of DM further exacerbates this discrepancy, leading these models to deviate more substantially from observed mass limits as $k_f^{\rm DM}$ increases. This indicates that these softer EOS models are less compatible with scenarios involving significant DM content, as they are unable to support NS masses within observational constraints even at baseline conditions without DM. 

Figure \ref{fig:figure13} examines the tidal deformability of a canonical NS ($\Lambda_{1.4M_{\odot}}$) as influenced by increasing $k_f^{\rm DM}$  for the same set of DM masses. All RMF models, including softer ones like FSUGold and IU-FSU, show a general trend of decreasing tidal deformability with increasing $k_f^{\rm DM}$. Interestingly, despite their limitations in supporting NS masses within the observational constraints for PSR J0740+6620, softer models like FSUGold and IU-FSU are still able to provide valuable constraints on tidal deformability. For higher DM masses, these models maintain tidal deformability within the observational range set by GW170817. Figure \ref{fig:figure14} shifts focus to the NICER constraints on the radius of a 1.108 $M_{\odot}$ NS for PSR J0030+0451, where observational data requires that the radius be no less than 10.13~km. Similar to the trends observed in maximum mass and tidal deformability, all RMF models show a decreasing radius with increasing $k_f^{\rm DM}$. However, even softer RMF models, which are generally quicker to fall below the maximum mass thresholds, are able to sustain radii above 10.13~km up to a certain point of $k_f^{\rm DM}$ for higher DM masses. This reveals that, while these models may struggle with mass constraints, they can still offer viable scenarios where the radius remains within acceptable limits, contributing to the broader understanding of DM effects on NS structure.
\begin{figure}[tbp]
    \centering
    \includegraphics[width=\columnwidth]{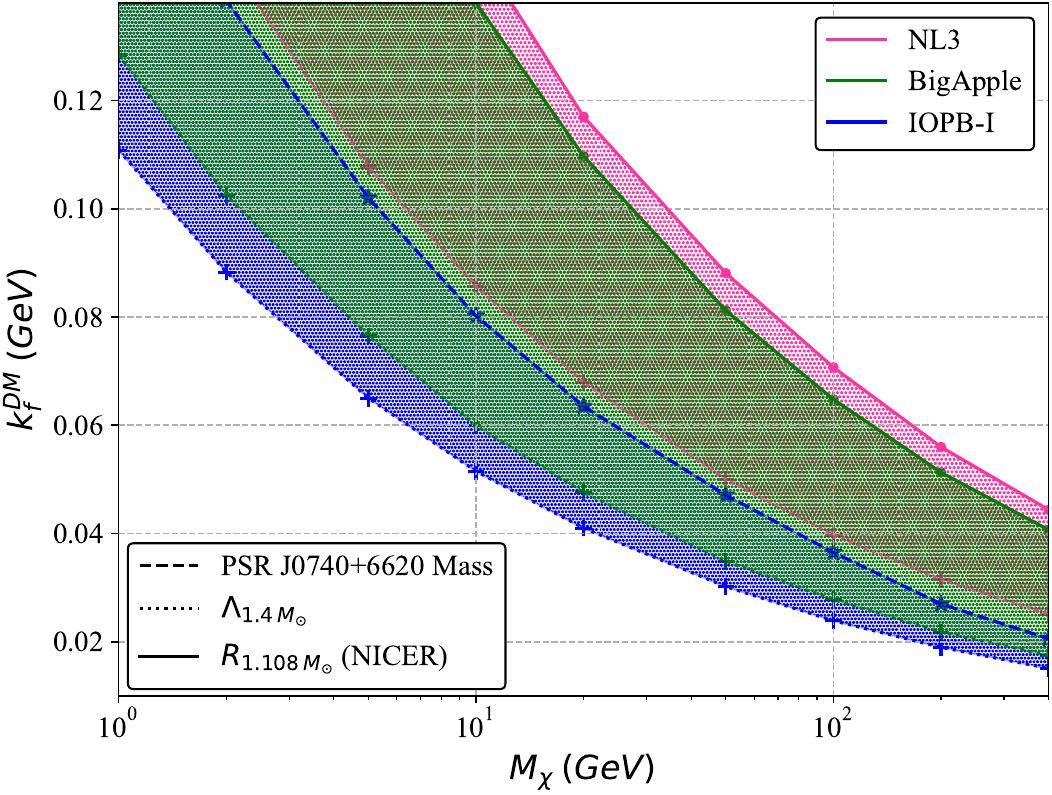} 
    \caption{Comparing the allowed regions for the NL3, BigApple, and IOPB-I parameter sets. The region except for the shaded regions should be excluded from the astronomical constraints on NS mass and radius. Considering that the NL3 parameter set is already excluded from the terrestrial experiments, the allowed region may become more severe. Or, we may say that the overlap region between the allowed regions with the BigApple and IOPB-I parameter sets is the ($M_\chi$, $k_f^{\rm DM}$) parameter space, which is allowed from the astronomical observations independently of the EOSs.}
    \label{fig:figure16}
\end{figure}

Figure \ref{fig:figure15} illustrates the constraints on DM parameters within NSs based on different observational constraints and for two distinct RMF models known for their relatively stiffer EOS, i.e., NL3 and BigApple. The analysis in Fig.~\ref{fig:figure15}, akin to the previously discussed Fig.~\ref{fig:figure11}, is structured around the influence of DM on NS properties — specifically the maximum mass, tidal deformability of canonical stars, and the radius of NSs as per NICER's observational constraints. Due to the inherently stiffer nature of NL3 and BigApple, these models tend to support larger NS masses at higher $k_f^{\rm DM}$ values even for higher $M_\chi$, aligning well with the observational mass limits of PSR J0740+6620 (Fig. \ref{fig:figure12}). This suggests that stiffer EOS can accommodate higher DM influences without compromising the mass stability of NSs. The analysis of tidal deformability, $\Lambda_{1.4M_{\odot}}$, from Fig. \ref{fig:figure13} further demonstrates that these stiffer models manage to maintain deformability within the empirical limits derived from GW170817 up to relatively high values of $k_f^{\rm DM}$. Each plot within Fig. \ref{fig:figure15} integrates results from a series of analyses (i.e. combining the constraints from maximum mass, canonical tidal deformability, and NICER radius observations) to outline permissible regions of DM mass and DM Fermi momentum that align with astrophysical observations.

The analysis in Fig.~\ref{fig:figure15} for the NL3 and BigApple parameter sets, alongside previously discussed results for IOPB-I, highlights a significant observation regarding the defining constraints on the upper limits of $k_f^{\rm DM}$ for a given $M_\chi$. This boundary is crucial as it determines the extent to which DM can influence NS properties without conflicting with empirical observations. For the stiffer RMF models such as NL3 and BigApple, the right boundary of the DM parameter space representing the maximum permissible $k_f^{\rm DM}$ for a given DM mass - is predominantly defined by NICER observations of the radius of a 1.108 $M_{\odot}$ NS from PSR J0030+0451. The ability of these models to support larger NS radii at higher $k_f^{\rm DM}$ and $M_\chi$ reflects their inherent robustness in maintaining structural integrity under DM influences. Conversely, for the IOPB-I model, which exhibits a slightly softer EOS compared to NL3 and BigApple, the defining boundary of the upper limit for $k_f^{\rm DM}$ is set by the mass limit of PSR J0740+6620.  The IOPB-I model, being less stiff, reaches this mass limit at a lower $k_f^{\rm DM}$ compared to how the radius constraint limits stiffer models, indicating a tighter constraint on DM parameters due to the softer EOS properties. This variation in how different RMF models are constrained by observational data underscores the critical role of EOS stiffness in theoretical predictions of NS properties in the presence of DM. Stiffer models like NL3 and BigApple can accommodate a broader range of DM parameters while still conforming to the radius constraints set by NICER. In contrast, the softer IOPB-I model is more restricted by mass constraints from PSR J0740+6620, demonstrating a more significant decrease in maximum mass with DM incorporation, which limits the permissible $k_f^{\rm DM}$ more stringently.

The final plot in this section synthesizes the allowed regions for DM parameter space across the above-discussed RMF parameter sets. Figure \ref{fig:figure16} compares the allowed regions for the DM parameters $M_\chi$ and $k_f^{\rm DM}$ for the NL3, BigApple, and IOPB-I parameter sets, each shaded differently to distinguish their respective permissible spaces. The overlay of these regions provides a visual representation of where the constraints from different models intersect, offering a composite view of the possible DM parameters that can exist without conflicting with empirical observations of NS mass and radius constraints. Notably, the NL3 parameter set, while offering some overlap with the regions permitted by BigApple, is also subject to exclusions based on terrestrial experiments, which have previously challenged its broader applicability. This adds a layer of complexity to interpreting the allowed parameter space, as the exclusion of regions supported by NL3 due to terrestrial constraints suggests a more stringent limitation on the DM parameters that are considered feasible. The intersection or overlap of the allowed regions for BigApple and IOPB-I sets forth a critical DM parameter space that is consistent with astronomical observations independent of specific EOS assumptions. This overlap highlights the robustness of the DM parameter constraints, suggesting that despite the variations in the stiffness and other properties of the nuclear matter EOS, there are fundamental limits on DM properties that remain consistent across different theoretical backgrounds. Meanwhile, individual RMF models exhibit unique sensitivities to DM influences—reflected in how each model's permissible DM parameter space is shaped—the overlap of allowed regions suggests underlying consistencies in DM constraints that transcend specific EOS formulations. The exclusion of certain DM parameters based on terrestrial experiments, particularly relevant to the NL3 model, combined with the astronomical constraints, highlights the importance of integrating multi-modal scientific observations and experiments in delineating the properties of DM within astrophysical observational efforts.

\subsection{Relation between $\Lambda$ and $M/R$}
\label{sec:3c}

In this subsection, we explore the universal relation between the dimensionless tidal deformability ($\Lambda$) and the compactness ($M/R$) of NSs. This relationship is known to hold independently of the theoretical framework, parameter sets (EOS), structure, chemical composition or any other factors affecting NSs. We aim to investigate whether this universality persists even in the presence of dark matter DM components for the adopted DM model in this study \cite{PhysRevD.96.083004}. Dimensionless tidal deformability, $\Lambda$, is a measure that quantifies the extent to which a NS deforms in response to the tidal field of its companion in a binary system. Mathematically, $\Lambda$ is expressed as \cite{PhysRevD.81.123016}:
\begin{equation}
    \Lambda = \frac{2}{3} k_{2} \left(\frac{R}{M}\right)^{5},
    \label{eq:Lamda_MR_original}
\end{equation}
where $R$ and $M$ are the radius and mass of the star; and $k_{2}$ is defined as the quadrupole dimensionless tidal love number which depends on the structure of the star \cite{PhysRevD.82.024016}.
\begin{figure}[tbp]
    \centering
    \includegraphics[width=\columnwidth]{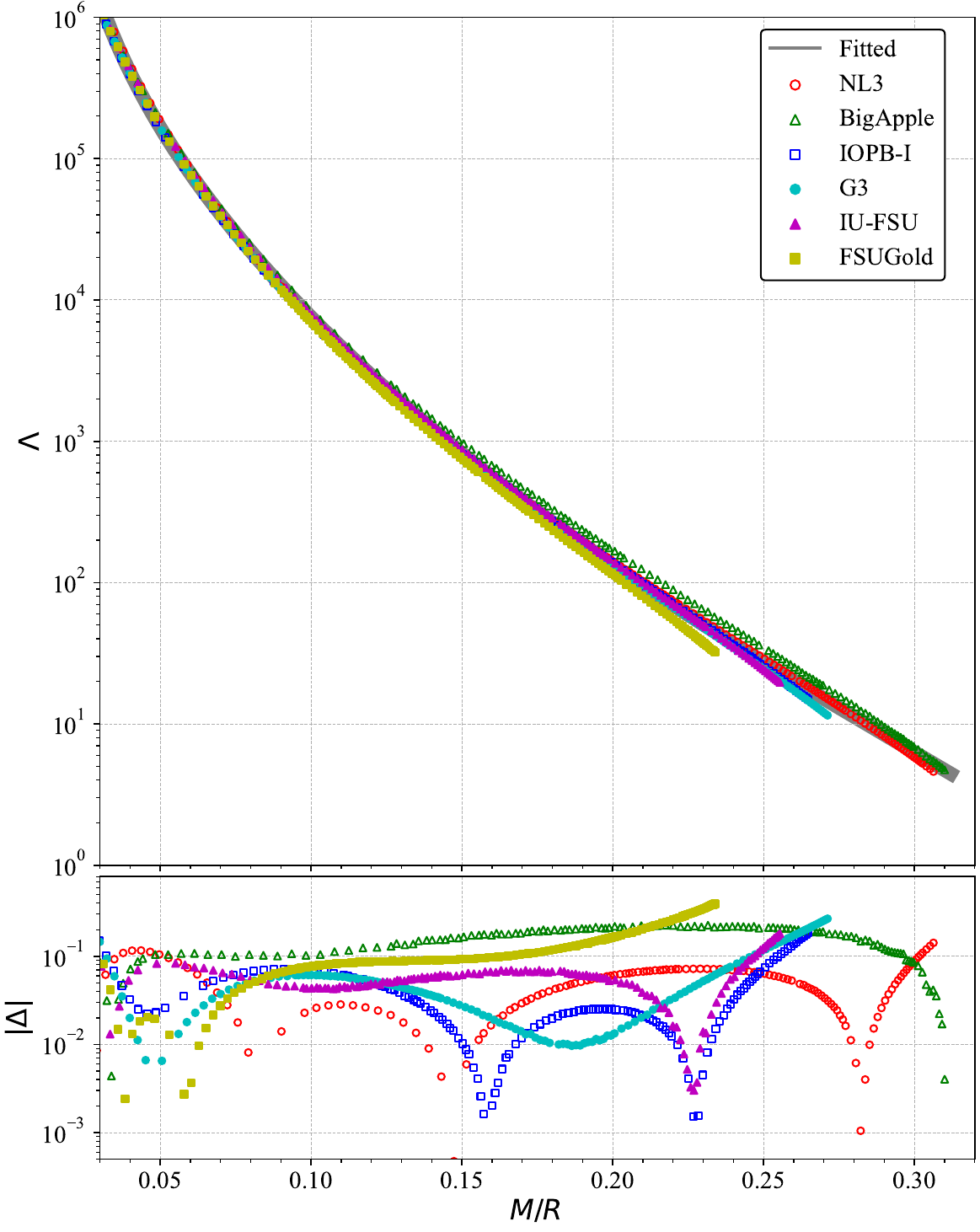} 
    \caption{In the top panel, the dimensionless tidal deformability, $\Lambda$, is shown as a function of the stellar compactness, $M/R$, for various EOS models in the absence of DM. The thick-solid line denotes the fitting line given by Eq.~(\ref{eq:Lamda_MR}), which is a kind of universal relation. In the bottom panel, the absolute value of the relative deviation from the estimation with the fitting line, given by Eq.~(\ref{eq:Delta}), is shown.}
    \label{fig:figure17}
\end{figure}

In Fig.~\ref{fig:figure17}, the dimensionless tidal deformability is plotted against the compactness for six different RMF parameter sets (NL3, BigApple, IOPB-I, G3, IU-FSU, FSUGold), all considering pure RMF EOSs without any DM inclusion. Through these models, we establish an empirical relationship between $\Lambda$ and $M/R$ that demonstrates a universal behavior across various EOS, indicative of the underlying physics that is independent of the specific models exploring NS matter and could potentially be used to infer properties of NSs across different observational scenarios and theoretical models. The figure includes a fitted line that represents this empirical relationship derived from the data, highlighting a consistent pattern across different RMF models and defined by the equation:
\begin{equation}
    \log_{10} \Lambda =  \frac{0.1641}{X} + 5.7791 - 5.3095 X + 1.9191 X^{2} - 0.4275 X^{3},
    \label{eq:Lamda_MR}
\end{equation}
where $X$ is scaled as $X \equiv (M/R)/0.2$. The bottom panel of the figure displays the absolute values of the relative errors for each RMF parameter set with respect to the empirical fitting formula. This error analysis evaluates the consistency and reliability of the empirical relationship across different theoretical models. The small and consistent error margins underscore the robustness of the fitted relationship, suggesting its potential applicability as a universal tool for predicting NS properties across a broad spectrum of scenarios.
\begin{figure*}[tbp]
    \centering
    \includegraphics[width=\textwidth]{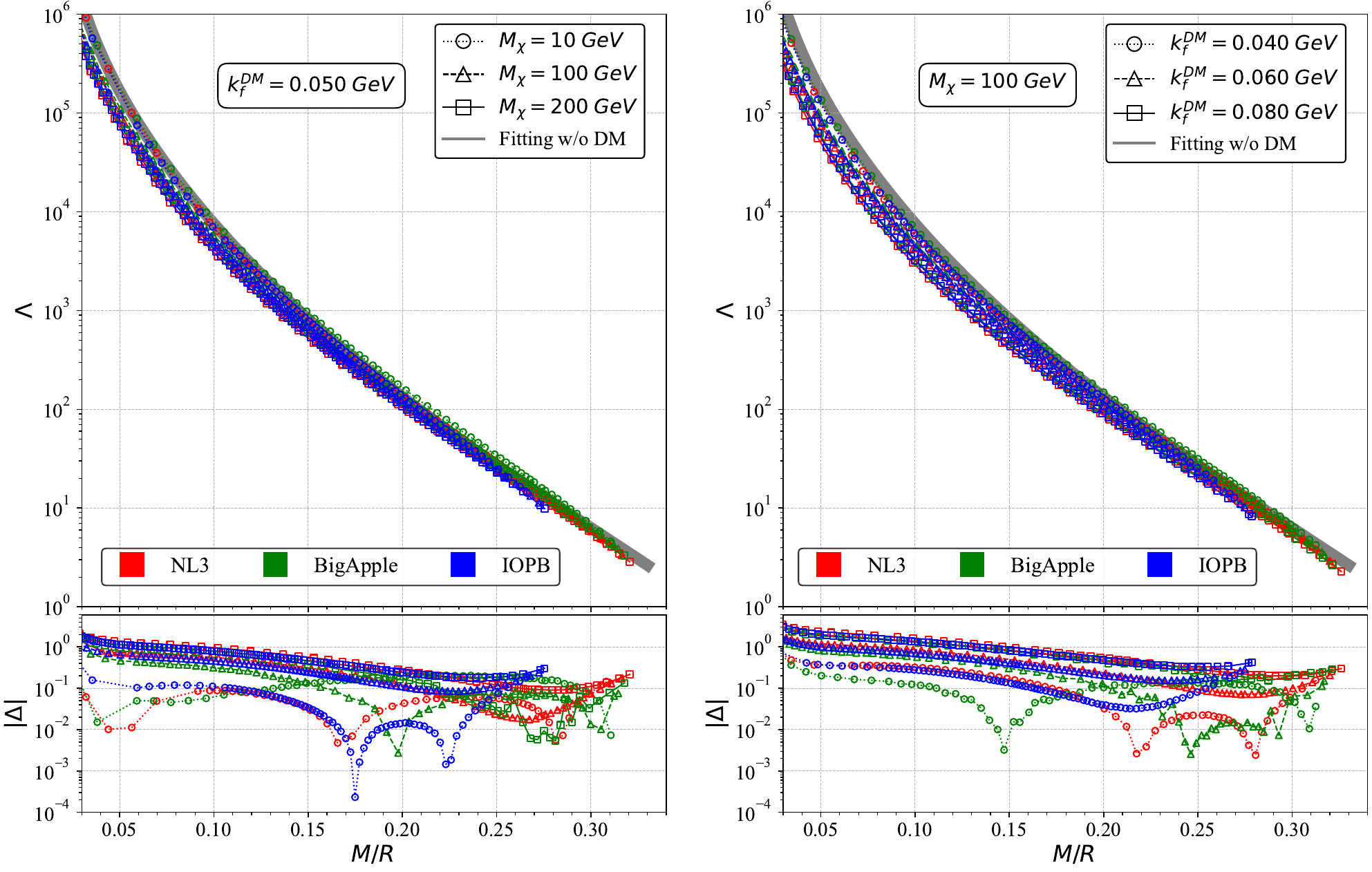} 
    \caption{The values of $\Lambda$ for the DM admixed NSs constructed with various EOSs are shown as a function of $M/R$. The left panel corresponds to the results by varying $M_\chi$ with the fixed value of $k_f^{\rm DM}=0.05$ GeV, while the right panel is by varying $k_f^{\rm DM}$ with the fixed value of $M_\chi=100$ GeV. In both panels, the universal relation obtained in the NS models without DM given by Eq.~(\ref{eq:Lamda_MR}) is shown with the thick-solid line. In the bottom panels, the absolute value of the relative deviation from the solid line is shown.}
    \label{fig:figure18}
\end{figure*}

The next figure in this subsection (Fig. \ref{fig:figure18}) presents a critical examination of the above-established empirical relationship between $\Lambda$ and $M/R$ of NSs, extending the analysis to include scenarios with varying DM parameters. This analysis aims to test the universality of the relationship in the presence of DM and to evaluate the need for potential modifications to the current DM-admixed model in light of this universal relation. The left plot in Fig. \ref{fig:figure18} depicts $\Lambda$ as a function of $M/R$ for a fixed DM Fermi momentum ($k_f^{\rm DM}=0.05$) across several DM masses (i.e., 10, 100, and 200 GeV). The empirical relationship derived without DM is included for comparison. Observations show that as DM mass increases, the deviations from the empirical relationship - or, in principal deviation from the universal relation - also increase, particularly at lower $M/R$ values, indicating a significant impact of DM on the tidal deformability of less compact NSs. In the right side plot of Fig. \ref{fig:figure18}, $\Lambda$ is evaluated against $M/R$ for a fixed DM mass ($M_{\chi} = 100$ GeV) while varying $k_f^{\rm DM}$ (i.e., 0.040, 0.060, and 0.080 GeV). This plot further supports the findings from the first analysis, showing increased deviations from the established empirical relationship as the value of $k_f^{\rm DM}$ increases. The results clearly demonstrate that the introduction of DM parameters significantly influences the tidal deformability of NSs, resulting in notable deviations from the previously established universal empirical relationship. These deviations are more pronounced for NSs with lower compactness and increase with higher DM mass or Fermi momentum at least within our adopted DM model. This could be attributed to the increased relative influence of DM on the overall mass-energy balance within less compact stars, potentially leading to significant changes in their deformation properties under tidal forces. Further research should focus on refining the theoretical framework to include the effects of DM that can accurately predict NS properties across a broader range of astrophysical conditions. This will enhance the predictive power of models used in gravitational wave astronomy and aid in the interpretation of signals from NS mergers, providing deeper insights into the nature of dense matter and dark matter in the universe. Subsequent research will explore these dimensions.

\section{Conclusion}
\label{sec:4}
In this study, we comprehensively explored the influence of DM on NS properties within the framework of relativistic mean-field theory.  By incorporating a DM model, we examined how DM parameters, such as the DM particle's mass and Fermi momentum, impact the nuclear saturation properties, the EOS, and the mass-radius relationship of NSs. By integrating observational constraints from PSR J0740+6620, NICER's analysis of PSR J0030+0451, and GW data from GW170817, we have delineated a multidimensional parameter space for DM that is consistent with current astrophysical observations. Our analysis also extended to the universal relation between dimensionless tidal deformability and compactness of NSs, assessing whether this relationship holds in the presence of DM. Key insights from this work include the identification of critical DM Fermi momenta and masses that significantly alter the nuclear properties and structural configurations of NSs. 

Our findings reveal that the inclusion of DM significantly alters the nuclear saturation properties. As the DM Fermi momentum increases, the nuclear saturation density also increases, leading to higher incompressibility, $K_{0}$, values. However, this increase in $K_{0}$
does not necessarily indicate a stiffer EOS in the traditional sense due to the shift in saturation density. The symmetry energy $J_{0}$ and slope parameter $L_{\text{sym},0}$ also increase with higher DM Fermi momentum and mass, suggesting more asymmetric nuclear matter and a steeper density dependence of the symmetry energy.

We derived constraints on DM Fermi momentum and mass using observational data from PSR J0740+6620, GW170817, and NICER measurements of PSR J0030+0451. Our analysis revealed that the maximum mass constraint from PSR J0740+6620 imposes the most stringent limits on DM parameters, particularly for softer EOS models like IOPB-I. The tidal deformability constraint from GW170817 and the radius constraint from NICER also significantly restrict the DM parameter space, particularly for stiffer EOS models like NL3 and BigApple. By overlaying these constraints, we identified a permissible region in the DM parameter space that is consistent with all observational data. This permissible region of DM parameters depends significantly on the chosen EOS. Stiffer EOS models like NL3 and BigApple can support higher maximum masses and larger radii even with higher DM content, while softer models like IOPB-I are more restricted by observational constraints. This variation underscores the critical role of EOS stiffness in theoretical predictions of NS properties in the presence of DM. The intersection of constraints from different observational sources and theoretical models has allowed us to delineate a robust set of permissible DM parameters, highlighting the importance of integrating multi-modal scientific observations to delineate the properties of DM within NSs. Moreover, this study challenges the assumption of universality in the presence of DM, showing that DM parameters could lead to deviations from the established empirical relationship, particularly for less compact NSs. This suggests a need to refine theoretical models to better account for DM effects, enhancing the predictive power and accuracy of NS models used in multimessenger astronomy.

Future work will focus on further refining these models to accommodate a broader range of DM effects and exploring additional observational data to tighten the constraints on DM properties. 
We acknowledge that the potential decay and/or annihilation processes of DM particles could be an important aspect to consider in this direction. While our current study does not address this, it is crucial to explore how stellar properties might evolve in light of such processes using the currently adopted DM model. This direction could reveal new insights into the interaction between DM and nuclear matter, further refining our theoretical framework.
Additionally, experimental investigations should consider the potential influence of DM on nuclear properties to provide a more complete understanding of nuclear matter in the presence of DM.

\begin{acknowledgments}
    \noindent This work is supported in part by Japan Society for the Promotion of Science (JSPS) KAKENHI Grant Numbers JP23K20848 and JP24KF0090, and by FY2023 RIKEN Incentive Research Project.
\end{acknowledgments}

\bibliographystyle{apsrev4-2}
\bibliography{ref.bib}

\end{document}